\documentclass[12pt,preprint]{aastex}

\usepackage{subscript}
\usepackage{hyperref}

\newcommand{\solar}{$_{\sun}$}
\newcommand{\te}{$T_{e}$}

\newcommand{\hiir}{\ion{H}{2}\\ region}

\newcommand{\cldy}{{\tt Cloudy}}
\newcommand\nd{\nodata}

\shorttitle{Carbon Abundances in WR Galaxies}
\shortauthors{Pe\~na-Guerrero et al.}

\received{}
\begin{document}

\title{Carbon Abundances in Starburst Galaxies of the Local Universe}

\author{Mar\'ia A. Pe\~na-Guerrero\footnotemark[1]}
%\affil{3700 San Martin Drive,
%Baltimore, MD, 21218, USA}
\email{pena@stsci.edu}

\author{Claus Leitherer\footnotemark[1]}
%\affil{3700 San Martin Drive,
%Baltimore, MD, 21218, USA}
\email{leitherer@stsci.edu}

\author{Selma de Mink\footnotemark[2]}
%\affil{Anton Pannekoek Institute for Astronomy. Science Park 904. P.O. Box 94249. 1090 GE, Amsterdam, The Netherlands}
\email{S.E.deMink@uva.nl}

\author{Aida Wofford\footnotemark[3]}
%\affil{Instituto de Astronom\'ia, UNAM, Ensenada, CP 22860, Baja California, Mexico}
\email{awofford@astrosen.unam.mx}

\and

\author{Lisa Kewley\footnotemark[4]}
%\affil{Research School of Astronomy and Astrophysics, Australian National University, Cotter Rd., Weston ACT 2611, Australia}
\email{lisa.kewley@anu.edu.au}

\footnotetext[1]{Space Telescope Science Institute. 3700 San Martin Drive, Baltimore, MD, 21218, USA}
\footnotetext[2]{Anton Pannekoek Institute for Astronomy. Science Park 904. P.O. Box 94249. 1090 GE, Amsterdam, The Netherlands}
\footnotetext[3]{Instituto de Astronom\'ia, UNAM, Ensenada, CP 22860, Baja California, Mexico}
\footnotetext[4]{Research School of Astronomy and Astrophysics, Australian National University, Cotter Rd., Weston ACT 2611, Australia}

\begin{abstract}
The cosmological origin of carbon, the fourth most abundant element in the Universe, is not well known and matter of heavy debate. We investigate the behavior of C/O to O/H in order to constrain the production mechanism of carbon. We measured emission-line intensities in a spectral range from 1600 to 10000 \AA\ on Space Telescope Imaging Spectrograph (STIS) long-slit spectra of 18 starburst galaxies in the local Universe. We determined chemical abundances through traditional nebular analysis and {we used a Markov Chain Monte Carlo (MCMC) method to determine where our carbon and oxygen abundances lie in the parameter space. We conclude that our C and O abundance measurements are sensible}. We analyzed the behavior of our sample in the [C/O] vs. [O/H] diagram with respect to other objects such as DLAs, neutral ISM measurements, and disk and halo stars, finding that each type of object seems to be located in a specific region of the diagram. Our sample shows a steeper C/O vs. O/H slope with respect to other samples, suggesting that massive stars contribute more to the production of C than N at higher metallicities, only for objects where massive stars are numerous; otherwise intermediate-mass stars dominate the C and N production.
\end {abstract}


\section{Introduction}\label{intro}
Carbon is the fourth most abundant element in the Universe as well as one of the key ingredients for life as we know it. It is a ubiquitous element in the interstellar medium (ISM): most molecules in the ISM are C-bearers (with CO being the most abundant molecule), and carbonaceous dust particles represent an important fraction of the ISM dust composition \citep[e.g.][and references therein]{gar95, dwe98, dwe05, rom14, zhu14}. Carbon also has an important role in regulating the temperature of the ISM: it contributes to the heating of the interstellar gas because it is the main supplier of free electrons in diffuse clouds, and it also contributes to the cooling of the warm interstellar gas through the emission of 158 $\mu$m \ion{C}{2} \citep{sta91, gul15}. Despite its relevance, there are only a few nebular carbon abundance determination studies because its brightest collisionally excited lines (CELs) are [\ion{C}{3}] 1907 and \ion{C}{3}] 1909 \AA \footnote{For simplicity we will refer to [\ion{C}{3}] 1907 and \ion{C}{3}] 1909 \AA\ as \ion{C}{3}] 1907+09 \AA.} and [\ion{C}{2}] 2326$\,$\AA\ in the UV \citep[e.g.][]{duf82, gar95, gar99, pen12a, sta14a}, and [\ion{C}{2}] 158 $\mu$m in the far-IR \citep[e.g.][]{tie85, sta91, mad97, gar04, can15}. The brightest carbon recombination line (RL), \ion{C}{2} 4267 \AA, is in the optical and it is only detectable and measurable in bright Galactic and extragalactic nebulae \citep[e.g.][]{est02, est05, lop07, est09, est14, gar05, pen12a}. RLs are intrinsically very weak, hence large resolving power is required to accurately measure such lines.

It is generally agreed that massive stars (M $>$ 8M\solar) synthesize most of the oxygen whereas carbon is synthesized by both low- and intermediate-mass stars as well as by massive stars \citep[e.g.][]{cla83, cow95, hen00}. In low-metallicity environments carbon is thought to be ``primary". Primary nucleosynthesis is defined as all those nuclei that can be produced from the initial H and He present in the star, it is independent of the initial stellar metallicity (e.g. $^{12}$C, $^{16}$O,  $^{20}$Ne,  $^{24}$Mg,  $^{28}$Si); secondary nucleosynthesis is defined as those nuclei that can be produced using preexisting nuclei from previous stellar generations, hence it is dependent on the initial stellar metallicity (e.g. $^{17}$O, $^{18}$O) \citep{mey05, lug12}. A particularly interesting element is $^{14}$N, which is produced by both primary and secondary nucleosynthesis \citep[e.g.][and references therein]{vin16}, but this element will not be discussed in this paper. Secondary production is common to stars of all masses \citep{mat86}. In high-metallicity environments, C, N, and O are synthesized during the CNO cycle as catalysts to produce He in both intermediate-mass and massive stars; after the hydrogen-burning phase (through proton-proton chain or CNO cycle), carbon and oxygen are byproducts of the triple-$\alpha$ process \citep{ren81}. C and O are then taken onto the surface of the star through the dredge-up process. As the mass of the star increases, He and C are removed from the star before forming O. 

The debate on the mass of the stars that contribute the most to the production of carbon is complicated due to the existence of different yields in stellar chemical evolution models \citep{car05}. There are currently several uncertainties in the carbon yields of massive stars. If mass-loss rates depend on metallicity, the yields of C and O also depend on metallicity, with C increasing at the expense of O \citep{gar95}. 
The amount of carbon and oxygen ejected by a star is directly dependent on its mass and on its metallicity. Both $^{16}$O and $^{12}$C are products of the triple-$\alpha$ process. Stellar evolution models predict that the mass of the least massive star capable of producing and ejecting new oxygen, is about 8 M\solar\ \citep{gar95}, less massive stars simply leave it in the core. In the case of carbon, the minimum stellar mass is predicted to be 2 to 8 M\solar\ in order to account for both massive stars and intermediate-mass through the dredge-up process during their AGB phase \citep{boy13}.

Previous work has shown that the yield of carbon varies with metallicity. \citet{car05} found that out of 11 Galactic chemical evolution models with different yields adopted for carbon, nitrogen, and oxygen, only two models fit the oxygen as well as the carbon gradient. These two models had carbon yields that increase with metallicity due to winds of massive stars, and decrease with metallicity due to winds of low- and intermediate-mass stars. \citet{fab09} determined [C/O] for 43 metal-pool halo stars, which are in reasonably good agreement with the results of \citet{car05}. \citet{boy13} suggested that a reduction in the carbon-rich to oxygen-rich AGB stars (respectively referred to as C and M stars) with metallicity is required in all modern TP-AGB models. This requirement comes from (i) the larger amount of carbon to be dredged-up to make the C/O$>$1 transition, and (ii) the third dredge-up starting later at higher luminosities and being less efficient at increasing metallicity. Note that a C star must have C/O$>$1, whereas an M star has C/O$<$1.
When the effects of rotation are included in stellar models, there is a very large increase in the yields of primary C, N, and O at very low metallicity \citep{mey02}. Furthermore, the majority of massive stars are suspected to be in close binaries and experience interaction with a companion \citep{san12}. The implications for the yields of massive stars have not yet been systematically studied. 

Carbon is crucial for the composition of interstellar dust. In order to obtain accurate ISM abundances, it is paramount to account for the presence of dust grains. Several studies have found that oxygen depletion in the Orion Nebula amounts to a correction in the total O/H of about 0.09 dex \citep{est98, est04, mes09, sim11}. A correction of 0.09 to 0.11 dex was suggested by \citet{pei10} for O/H of \ion{H}{2} regions, depending on the metallicity of the object. Most of the ISM dust can be broadly classified into either carbonaceous- or silicone-based, hence both types of dust affect our study of C/O. Dust formation can be broadly divided into two types of sources: (i) those that undergo quiescent mass-loss (e.g. W-R stars) and (ii) those that return their ejecta eruptively back into the ISM (e.g. Type Ia and Type II supernovae and asymptotic giant branch [AGB] stars). The type of dust does not necessarily depend on the formation source but rather on the C/O in the ejecta. For low and intermediate-mass stars (M$\;<8$M\solar), if C/O$\;>1$, all the oxygen is tied up in CO molecules and the newly formed dust grains will be carbon-rich; if C/O$\;<1$, the extra oxygen will combine with other elements to form silicate-based types of dust grains \citep{dwe98}. Massive stars will contribute to the dust grain production only with the coolest stars and according to the exposed material \citep{coe05}. 

Wolf-Rayet (W-R) galaxies are natural test beds for the study of stellar chemical evolution models for the enhancement of CNO elements in massive stars. W-R galaxies are a subset of the class of starburst galaxies or emission-line and \ion{H}{2} galaxies, whose integrated spectra present the ``starprint" of W-R stars, i.e. broad emission spectral features associated to W-R stars, being the main feature the broad He II $\lambda$4686 \AA\ emission line \citep[e.g.][and  L\'opez-S\'anchez \& Esteban 2008 among others]{ost82, kun85, con91}. W-R stars are chemically evolved end-stages of the most massive stars within a starburst region \citep{cro07}. W-R stars have very short lives (about 10$^5$ yr), hence they can only be detected in population when numerous. This implies that the starburst activity of W-R stars is dominant with respect to the lower-mass stars. Single stars with masses greater than 30 to 60 M\solar\ (depending on metallicity and rotation rate) become W-R \citep{mey94}. Therefore, the ``starprint" of W-R stars indicates a top-heavy initial mass function (IMF). L\'opez-S\'anchez \& Esteban 2008, 2009, 2010a, 2010b, and L\'opez-S\'anchez 2010 conducted the hereto most complete observational study of W-R galaxies. Their observations include ground-based optical spectra, deep broad- and narrow-band images, radio, and X-rays. From here on we will refer to the work of L\'opez-S\'anchez \& Esteban 2008, 2009, 2010a, and 2010b as LSE08, LSE09, LSE10a, LSE10b, respectively.

This study has two main motivations: (i) to determine the source of most of the carbon production (i.e. either massive or intermediate-mass stars), and (ii) to study the behavior of carbon as a function of chemical composition. This information will allow better constraints on stellar and galactic models of chemical evolution. To address these points, we used low-resolution Hubble Space Telescope (HST) Space Telescope Imaging Spectrograph (STIS) long-slit spectra of 18 local starburst galaxies. This work is divided into the following sections: sample selection is described in Section \ref{sampsel}, observation details and a general description of the sample are described in Section \ref{obs}, the observations and data reduction are presented in Section \ref{met}, the data analysis and methodology, including line flux analysis and reddening correction, a brief description of the Direct Method, the physical conditions of the ionized gas, {and the} chemical composition; the discussion and summary and conclusions are presented in Section \ref{disc} and \ref{conc}, respectively. {The Markov Chain Monte Carlo (MCMC) modeling of photoionized objects is presented in the Appendix.}

\section{Description of the Sample}\label{sampsel}
The sample for this paper was drawn from the W-R galaxies studied by \citet{lop08}. Their original sample includes 20 galaxies, however we removed two objects: NGC$\,$5253 since HST archival data already exist \citep{kob97}, and SBS$\,$1211+540 because its faintness required prohibitively long exposure times. This section describes each of the 18 objects of our W-R galaxy sample. The main properties of each object are presented in Table \ref{main_props}. We follow LSE08, LS09, LS10a,b and \citet{lop10} for most of the general information of each galaxy. We have verified that the regions of the objects observed in our HST STIS data {were} the same as those observed in the works of L\'opez-S\'anchez \& Esteban.

\subsection{Mrk 960}\label{mrk960}
Mrk 960 was catalogued as Haro 15 by \citet{har56} as a blue galaxy with emission lines. This object has been extensively studied in a wide range of wavelengths: i.e. in the UV \citep{kaz79, kin93, hec98}, in the optical \citep{cai01a, cai01b, lop08, lop09, lop10a, lop10b, fir11, hag12}, in the NIR \citep{coz01, dor13}, in the FIR \citep{cal94, cal95}, and in radio \citep{gor81, klei84, klei91}. Our STIS observations correspond to the center region, C, described in \citep{lop08}; in that work Mrk 960 is referred to as Haro 15.

\subsection{SBS 0218$+$003}\label{sbs0218}
SBS 0218$+$003 is included in the W-R galaxies catalogue of \citet{sch99}. It is the most distant object analyzed in this work as well as in \citet{lop08, lop09, lop10a, lop10b}, in which the object is referred to as UM 420. A note provided in the NASA/IPAC Extragalactic Database (NED) indicates that SBS 0218$+$003 is probably an \ion{H}{2} region in UGC 1809. L\'opez-S\'anchez et al. compared the spectra of SBS 0218$+$003 with that of UGC 1809 and concluded that the latter as an S0 spiral galaxy at redshift $z=0.0243$. By comparing the radial velocities of both objects they {concluded} that they are not physically related. In the work of \citet{lop08} SBS 0218$+$003 is referred to as UM 420.

\subsection{Mrk 1087}\label{mrk1087}
Mrk 1087 was classified by \citet{con91} as an emission-line galaxy without the broad emission line \ion{He}{2} 4686 \AA, and it was later classified as a luminous blue compact galaxy (BCG) within a group of dwarf objects in interaction by \citet{lop04b}. These authors argue that Mrk 1087 does not host an Active Galactic Nucleus, and that this galaxy and its dwarf companions should be considered a group of galaxies. According to \citet{lop04b}, the various filaments of Mrk 1087 and surrounding dwarf objects suggest that this could be a group in interaction. Such filaments were first reported by \citet{men00}. Our STIS observations of Mrk 1087 correspond to the center knot of \citet{lop04b} and \citet{lop08}.

\subsection{NGC 1741}\label{ngc1741}
NGC 1741 is the brightest member of the interacting group of galaxies HCG 31. According to \citet{lop04a}, the analysis of the kinematics of HCG 31 suggests that an almost simultaneous interaction involving several objects are taking place. In the nomenclature given by \citet{hic82}, NGC 1741 actually corresponds to HCG 31C, nonetheless since objects A and C are clearly interacting, the two objects can be considered a single entity called HCG 31 AC. A detailed analysis of these interacting galaxies in broad-band imaging and optical intermediate-resolution spectroscopy is presented in \citet{lop04a}. Our STIS observations correspond to the north east part of knot AC in \citet{lop04a} and \citet{lop08}.

\subsection{Mrk 5}\label{mrk5}
\citet{mar67} included Mrk 5 in his first list of galaxies with UV continua, later it was classified as an emission-line galaxy with a narrow \ion{He}{2} 4686 \AA\ in emission by \citet{con91}. It is usually classified as an \ion{H}{2} galaxy and/or a cometary-type Blue Compact Dwarf Galaxy (BCDG). It has an extensive, regular and elliptical envelope formed by old stars and it is a low metallicty object \citep{lop08}. Our STIS observations of Mrk 5 correspond to slit position INT-1 in the work of \citet{lop08}.

\subsection{Mrk 1199}\label{mrk1199}
Mrk 1199  is part of a group of interacting galaxies. The main body of the group is an Sb galaxy, which is interacting with an elliptical object located to the NE of the main galaxy. Mrk 1199 was classified as a W-R galaxy by \citet{sch99}. The [\ion{O}{3}] 4363 \AA\ line was reported as not detected in the work of \citet{izo98} and \citet{lop09}. We did not detect this line either, however we did observe a nebular \ion{He}{2} 4686 \AA\ emission line. Our STIS observations of Mrk 1199 correspond to slit position D in the work of \citet{lop08}.

\subsection{IRAS 08208$+$2816}\label{iras08208}
IRAS 08208$+$2816 is classified as an \ion{H}{2} galaxy and it is included in the W-R galaxies catalogue of \citet{sch99}. \citet{hua99} first reported both nebular and broad  \ion{He}{2} 4686 \AA\ emission lines, as well as the W-R blue and red bumps (at \ion{C}{3} 4650 \AA\ and \ion{C}{4} 5808 \AA, respectively), which suggest the presence of late type WN stars (WNL) and early type WC stars (WCE) populations in the galaxy. Our STIS observations of IRAS 08208$+$2816 correspond to knot C in \citet{lop08}.

\subsection{IRAS 08339$+$6517}\label{iras08339}
IRAS $08339+6517$ is a luminous infrared and Ly-$\alpha$ emitting starburst galaxy, catalogued as a W-R galaxy by \citet{lop06}. These authors presented a detailed study of deep broad-band optical images, narrow band H$\alpha$ CCD images, and optical intermediate- resolution spectra of IRAS $08339+6517$ and its dwarf companion. They concluded that the chemical composition of both galaxies is similar, and that these objects are most likely kinematically interacting. Our STIS observations of IRAS $08339+6517$ correspond to knot A in \citet{lop08}.

\subsection{SBS 0926$+$606A}\label{sbs0926}
SBS 0926$+$606A is one component of the pair of objects of SBS 0926$+$606, where component A is a BCDG and component B is a more elongated object north of object A and with no W-R features detected. \citet{izo94} first detected the narrow \ion{He}{2} 4686 \AA\ emission line in SBS$\;0926+606$A to measure the primordial helium abundance. The galaxy was later studied spectroscopically by several authors \citep[e.g.][]{izo97, per03, kni04}, and the properties of massive stars in this galaxy where studied by \citet{gus00}. Our STIS observations of SBS 0926$+$606 correspond to knot A in \citet{lop08}.

\subsection{Arp 252}\label{arp252}
Arp 252 is classified as an interacting pair of galaxies. Our STIS observations correspond to the brighter galaxy, ESO 566-8, which is the northern object. Most of the H$\alpha$ emission of the entire system (93\%) comes from the brighter galaxy \citep{lop08}. Our STIS observations of Arp 252 correspond to knot A (ESO 566-8) in \citet{lop08}.

\subsection{SBS 0948$+$532}\label{sbs0948}
SBS 0948$+$532 is an emission line galaxy and a BCDG included in the W-R galaxies catalogue of \citet{sch99}. \citet{izo94} first detected the \ion{He}{2} 4686 \AA\ emission line in this object. \citet{gus00} re-analyzed SBS 0948$+$532 and found the presence of WNL stars and tentative evidence of a red WR bump. 

\subsection{Tol 9}\label{tol9}
Tol 9 is the most metal-rich object in our sample. It is classified as an emission-line galaxy without the emission line \ion{He}{2} 4686 \AA, and as a W-R galaxy by \citet{sch99}. \citet{wam85} suggested that Tol 9 is interacting with a nearby object. Our STIS observations of Tol 9 correspond to slit position INT in the work of \citet{lop08}.

\subsection{SBS 1054$+$365}\label{sbs1054}
SBS 1054$+$365 is a BCDG included in the W-R galaxies catalogue of \citet{sch99}, and in the catalog of interacting galaxies of \citet{vor59, vor77} due to the detection of a nearby companion about 1 arcminute to the north. Our STIS observations focused on region C, which is the brightest knot; these observations correspond to the main component in the work of \citet{lop08}.

\subsection{POX 4}\label{POX4}
POX 4 is included in the catalogue of W-R galaxies of \citet{con91} as well as in the catalogue of \citet{sch99}. The broad \ion{He}{2} 4686 \AA\ emission line was first detected in POX 4 by \citet{kun85}. It is classified as a BCDG with its bright knot surrounded by three or four star-forming regions. \citet{lop10a} detected both broad \ion{He}{2} 4686 \AA\ and \ion{C}{4} 5808 \AA\ to determine the number of WNL and WCE stars. Our STIS observations of POX 4 correspond to the main component in the work of \citet{lop08}.

\subsection{SBS 1319$+$579}\label{sbs1319}
The \ion{He}{2} 4686 \AA\ emission line was first detected in the BCDG SBS 1319$+$579 by \citet{izo94}. \citet{sch99} included this object in their W-R galaxies catalogue, and later \citet{gus00} detected in it WNL and WCE populations. Our STIS observations correspond to knot A in \citet{lop08}, which is the brightest knot in SBS 1319$+$579. 

\subsection{SBS 1415$+$437}\label{sbs1415}
SBS 1415$+$437 is the most metal-poor object in our sample and it is one of the most metal-poor BCDGs known. From broad-band photometry, \citet{lop08} found that the brightest regions of the galaxy are blue, yet slightly redder when considering the flux from all the galaxy, suggesting the existence of a low-luminosity component dominated by older stellar populations. Our STIS observations of SBS 1415$+$437 correspond to knot A in the work of \citet{lop08}.

\subsection{Tol 1457$-$262}\label{tol1457}
Tol 1457$-$262 is classified as a pair of galaxies with significant star formation activity. W-R features have been detected in the western object by several authors \citep[e.g.][]{con96, pin99, lop08, lop09, lop10a, lop10b, est14}. Our STIS observations correspond to region B in \citet{lop08}. Tol 1457$-$262 is included in the W-R galaxies catalogue of \citet{sch99}. Our STIS observations of Tol 1457$-$262 correspond to knot A in the work of \citet{lop08}.

\subsection{III Zw 107}\label{iiizw107}
III Zw 107 is classified as an emission-line galaxy. It was named after the \it{Catalogue of Selected Compact Galaxies and of Post-Eruptive Galaxies}\upshape by \citet{zwi71}. 
Photometric studies have been performed on III Zw 107 by \citet{mol87} and \citet{cai01a, cai01b}. Spectrophotometric studies in the visual, X-ray, and radio on this galaxy have been performed by  \citet{lop08, lop09, lop10a, lop10b}.  \citet{kun85} included III Zw 107 in their catalogue of W-R galaxies. Our STIS observations of III Zw 107 correspond to knot A in the work of \citet{lop08}.

\section{Observations}\label{obs}

The sample was observed in Hubble Space Telescope (HST) program GO 12472 (PI: Leitherer), which uses STIS to perform co-spatial spectroscopy over the wavelength range of 1600 to 10,000 \AA. The final coordinates are given in Table \ref{tobs}, after target acquisition of the telescope. The distances higher {than} 20 Mpc were taken from the NED with the Hubble flow calculations assuming that $H_0=73$ km s$^{-1}$ Mpc$^{-1}$ and the Virgo, GA, Shapley model; closer distances were taken from the work of \citet{zha13}. 

\begin{figure}
\begin{center}
\includegraphics[angle=0,scale=0.5]{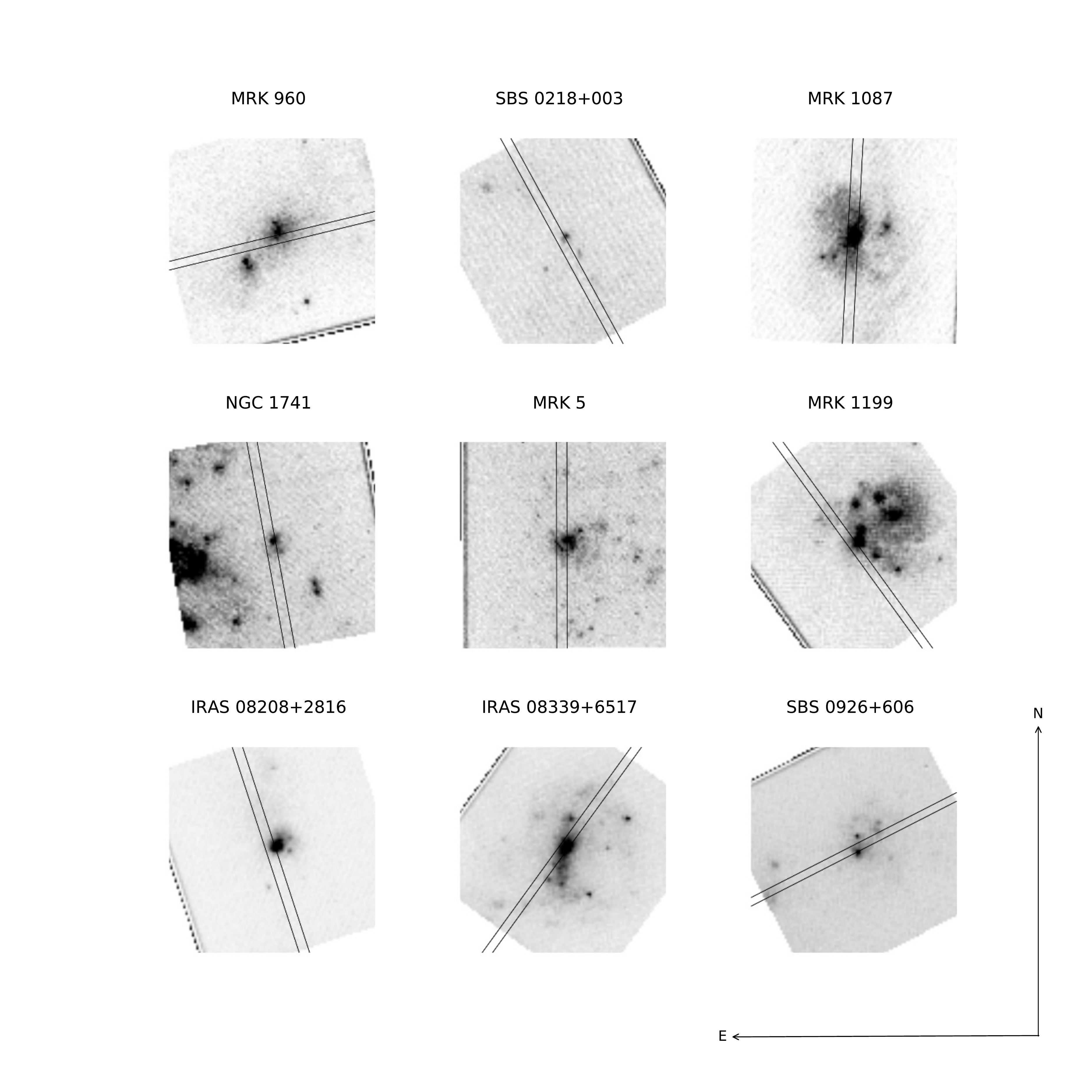}\\
\caption[tile1-eps-converted-to.pdf]{
Optical HST target acquisition images rotated north and showing slit positions for our STIS observations for the first half of the sample. Each image is $5\times5$ arcseconds or $100\times100$ pixels, hence the $0.2\times52$ slit used corresponds to 4 pixels. The angles presented in Table \ref{tobs} are with reference to north, indicated in the lower right side of the image; positive angles are to the east. The origin of the coordinate system is at the center of each acquisition image.
\label{tile1}}
\end{center}
\end{figure}

\begin{figure}
\begin{center}
\includegraphics[angle=0,scale=0.5]{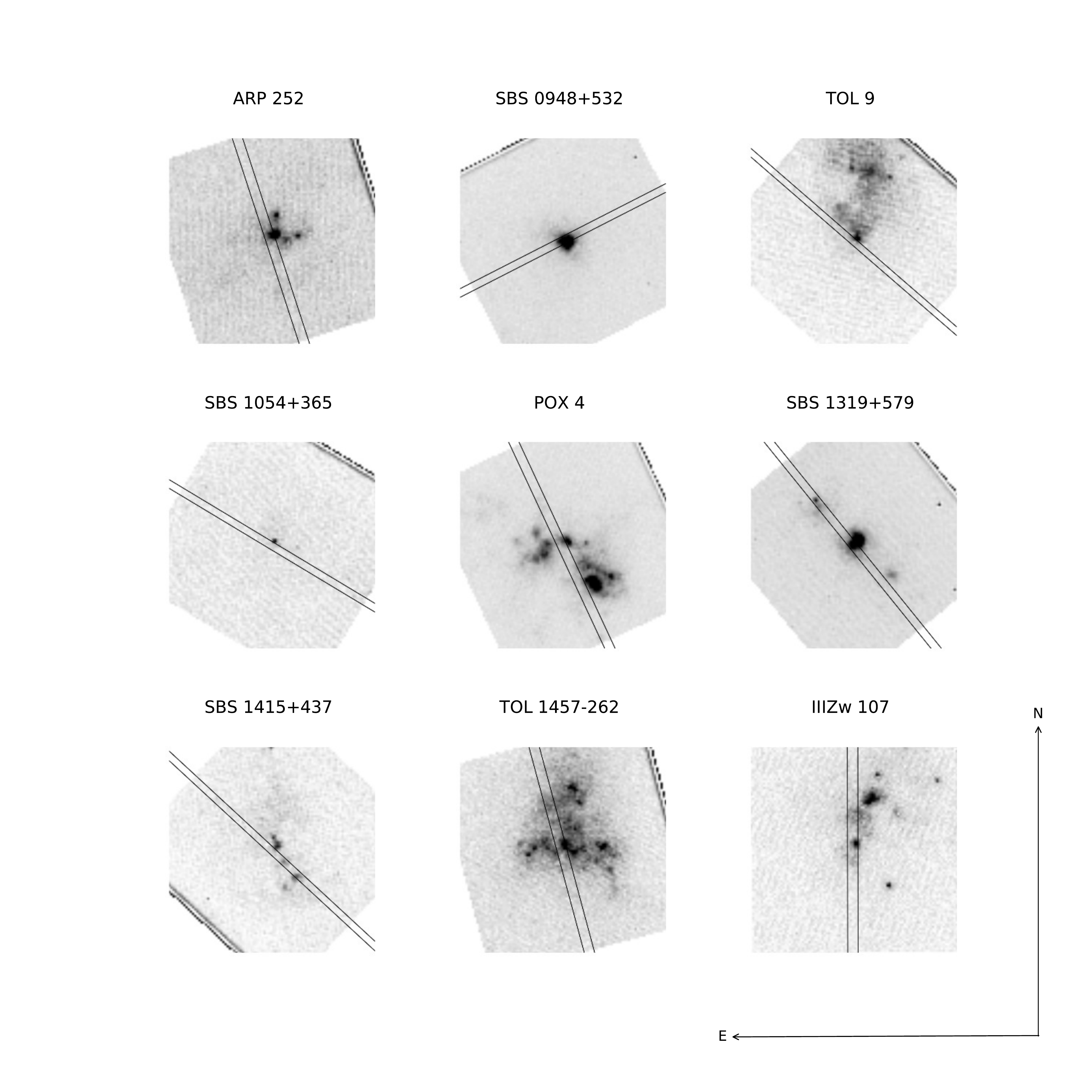}\\
\caption[tile2-eps-converted-to.pdf]{
Optical HST target acquisition images rotated north and showing slit positions for our STIS observations for the second half of the sample. Image and slit sizes are the same as in Figure \ref{tile1}.
\label{tile2}}
\end{center}
\end{figure}

The observation program was conducted between January 2012 and January 2014. We used the long-slit NUV-MAMA and CCD detectors with three gratings: G230L for the MAMA detector, {and} G430L and G750L for the CCD. The G230L grating has a spectral range from 1560-3180 \AA, and an average dispersion of 1.58 \AA/pixel; the G430L grating has a spectral range from 2900-5700 \AA, and an average dispersion of 2.73 \AA/pixel; and the G750L grating has a spectral range from 5240-10270 \AA, and an average dispersion of 4.92 \AA/pixel. All three gratings have {a resolving power, R, of about} 500.
The properties of the observations are described in Table \ref{tobs}. The spectra were taken with the $0.2''\times52''$ aperture, which is a good compromise between slit loss and spectral resolution at a R$\,\sim\,$500. Prior to taking the STIS spectra, we used the CCD detector to obtain a $5\times5$ arcsecond (or $100\times100$ pixels) target acquisition image. We used this image to check the acquisition of the STIS spectra. Figures \ref{tile1} and \ref{tile2} show the acquisition images. The slit positions are just as taken from the proposal. {The HST data used for this analysis can be downloaded from the Mikulski Archive for Space Telescopes (MAST) (\url{https://doi.org/10.17909/T96S3J}).}

\subsection{Data Reduction}\label{datared}
\begin{figure}
\begin{center}
\includegraphics[angle=0,scale=0.5]{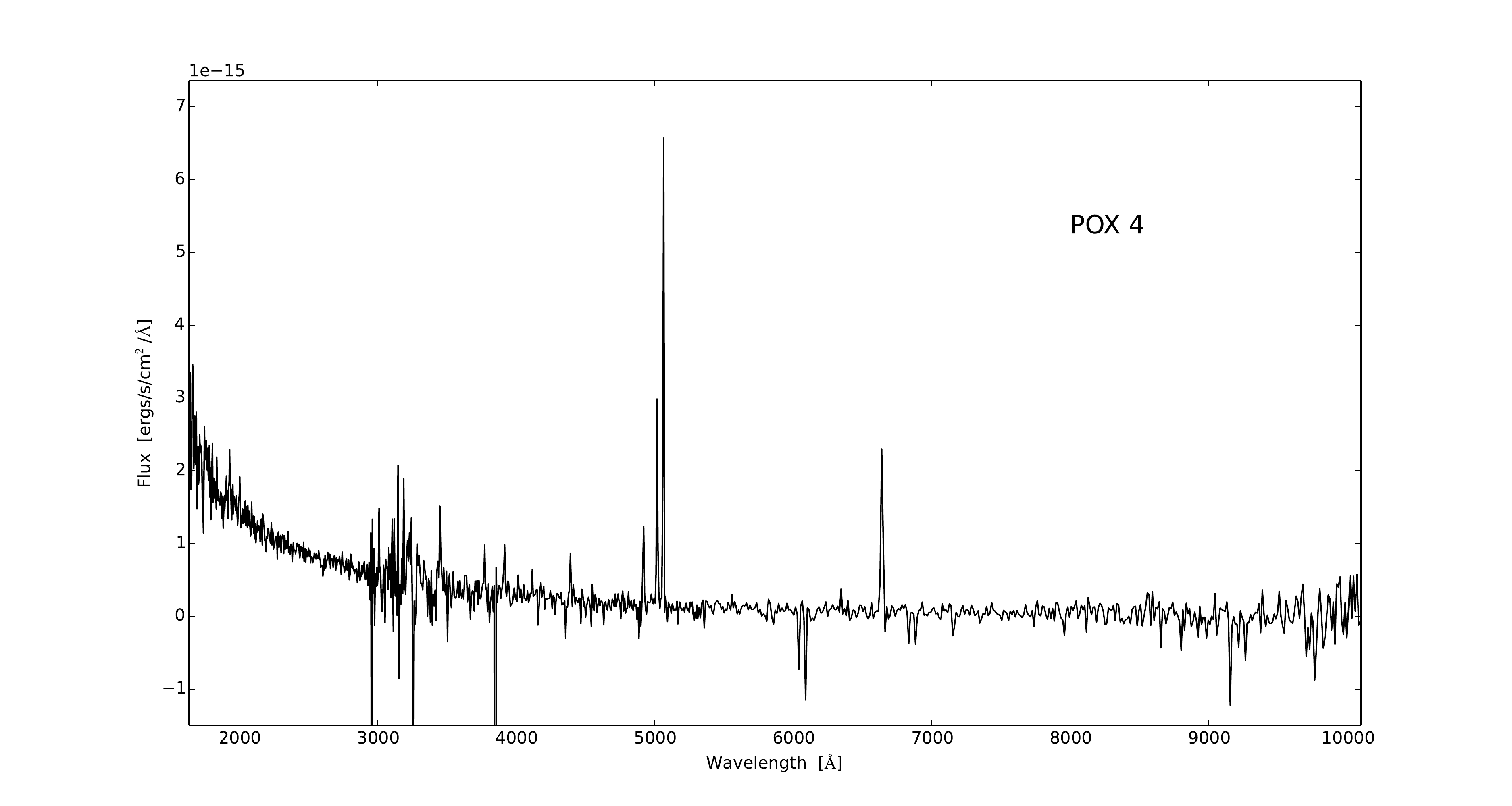}\\
\caption[pox4spec-eps-converted-to.pdf]{
Combined NUV, optical, and NIR HST STIS spectra of POX 4.
\label{pox4spec}}
\end{center}
\end{figure}

\begin{figure}
\begin{center}
\includegraphics[angle=0,scale=0.5]{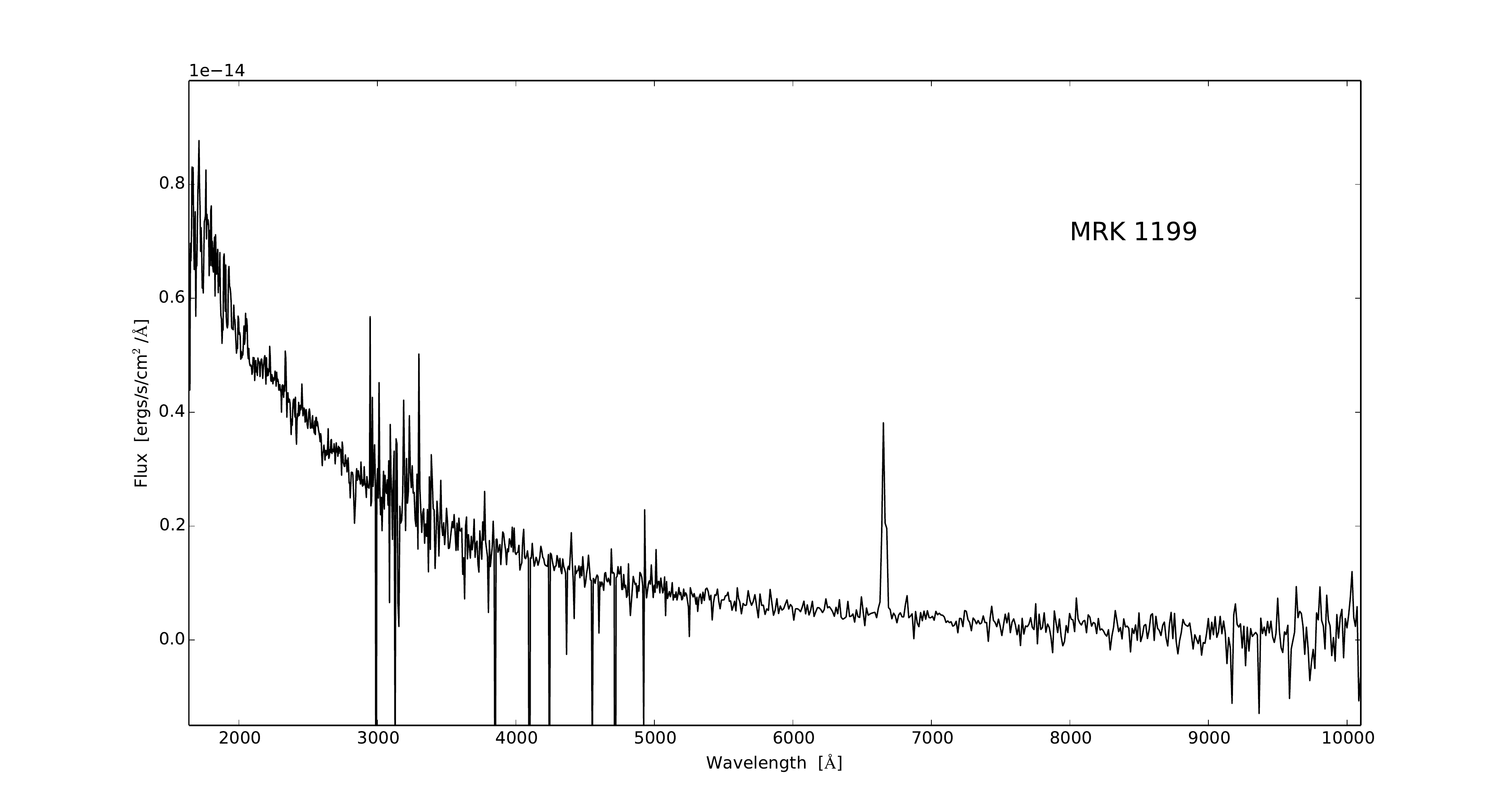}\\
\caption[mrk1199spec-eps-converted-to.pdf]{
Combined NUV, optical, and NIR HST STIS spectra of Mrk 1199.
\label{mrk1199spec}}
\end{center}
\end{figure}

The data were processed with the {\tt CALSTIS} pipeline \citep{bri15}, which includes the following steps: conversion from high-res to low-res pixels (MAMA), linearity correction, dark subtraction, cosmic-ray rejection (CCD), combination of cr-split images (CCD), flat fielding, geometric distortion correction, wavelength calibration, and photometric calibration. We extracted one-dimensional (1D) spectra of each object from the x2d (MAMA) and sx2 (CCD) files, which contain two-dimensional (2D) spectral images. Since some of our objects are very faint, we re-extracted all spectra with four different extraction windows: 11, 16, 21, and 30 pixels. The pipeline default for extended objects is 11 pixels for the MAMA spectra and 7 pixels for the CCD. For each NUV, optical, and near-IR spectra we chose the extraction window with the best signal-to-noise (S/N). The CCD spectra were also cleaned from cosmic rays using the Python module {\tt cosmics} based on Pieter van Dokkum's  {\tt L.A.Cosmic} \citep{van01}. We show the spectra of one of our best and worse S/N objects, POX 4 (Figure \ref{pox4spec}) and Mrk 1199 (Figure \ref{mrk1199spec}), respectively.

\section{Data Analysis and Methodology}\label{met}
We followed a traditional analysis using a two-zone approximation to define the temperature structure of the object{, and used it to calculate the oxygen abundances}. {We obtained the} C/H {abundances with} the method described in \citet{gar95}. {To determine if our carbon and oxygen abundances were sensible, we used a Markov chain Monte Carlo (MCMC) method to probe the parameter space}. This methodology is described in detail in the Appendix.

\subsection{Line Flux Analysis and Reddening Correction}
The procedure of line flux measurement was done with a Python code we developed. This code determines the stellar continuum, finds the emission lines from a catalogue we compiled, and measures the total flux. {This catalogue contains} typical emission and absorption lines observed in nebular spectra is composed as follows: lines from about 1150 to 2850 \AA\ were taken from \citet{lei11}, and lines from about 3200 to 10300 \AA\ were taken from \citet{pei03}. We measured all lines with a width at the continuum greater than 1.5 \AA. The continuum was determined by sigma clipping the strong emission lines and then finding the flux mode of the remainder signal. The flux of the emission lines was determined with a simple sum of flux over continuum routine. 

To obtain the reddening corrected intensities, we included PyNeb v.0.9.13 \citep{lur15} into our code. We used the extinction law by \citet{fit99} for the UV, and \citet{fit90} for the optical and NIR, both with with $R_V=3.1$. The Balmer and helium emission lines were corrected for underlying absorption, and the adopted EWs in absorption were taken from a stellar spectra template normalized to EW\textsubscript{abs} provided in Table 2 of \citet{pen12a}. This template was based on the low-metallicity instantaneous bursts models from \citet{gon99}, as well as additional models ran by M. Cervi\~no with the same code as Gonz\'alez-Delgado and collaborators. In order to find the values of C(H$\beta$) for this sample, since H$\alpha$ is blended with the [\ion{N}{2}] lines due to the low resolution (R$\sim$500), we did an iterative process using the expected theoretical value of H$\alpha$ according to \citet{sto95}, the measured flux of H$\beta$, H$\gamma$, and H$\delta$, and the deblend task in the Pyraf routine {\tt splot}.  The deblended intensity of H$\alpha$ is presented in Tables \ref{tlines1},  \ref{tlines2}, and \ref{tlines3}. 

The dereddend line intensities and final used values for C(H$\beta$) and EW\textsubscript{abs}, as well as the the equivalent widths for other important lines, are presented in Tables  \ref{tlines1},  \ref{tlines2}, and \ref{tlines3}. The structure is the same for all three tables: columns 1 and 2 are respectively the rest frame wavelength and the line identification (ID), column 3 is the reddening law used ($f_{\lambda}$ values), columns 4 through 9 show the dereddened line intensities relative to H$\beta$, with the standard assumption that $I$(H$\beta$)=100. The values of C(H$\beta$) we obtained agree within the errors with the values determined by \citet{lop09}.

To obtain uncertainties of the measured lines we used the estimated nominal spectroscopic accuracies for flux calibration in L mode given in the STIS Data Handbook, 2\%, 5\%, and 5\% for NUV, {optical}, and NIR, respectively. We assumed the uncertainties to be symmetric around the center wavelength. The contribution to the uncertainties due to the noise was estimated from the rms of the continuum adjacent to the emission line. The final adopted uncertainties were estimated using standard error propagation equations.

\subsection{Direct Method}\label{dirmet}
The so-called direct method assumes a homogeneous temperature structure throughout the whole volume of the object{, and then this temperature is used} to determine abundances of all available ions. This method generally adopts a two-{ionization} zone approximation, where the temperature of the high ionization zone can be represented the electron temperature of [\ion{O}{3}], \te\ [\ion{O}{3}], or \te\ [\ion{S}{3}] and the temperature of the low ionization zone can be represented by \te\ [\ion{O}{2}] or \te\ [\ion{N}{2}]. 

\subsection{Physical Conditions of the Ionized Gas}\label{physcon}
\begin{figure}
\begin{center}
\includegraphics[angle=0,scale=0.22]{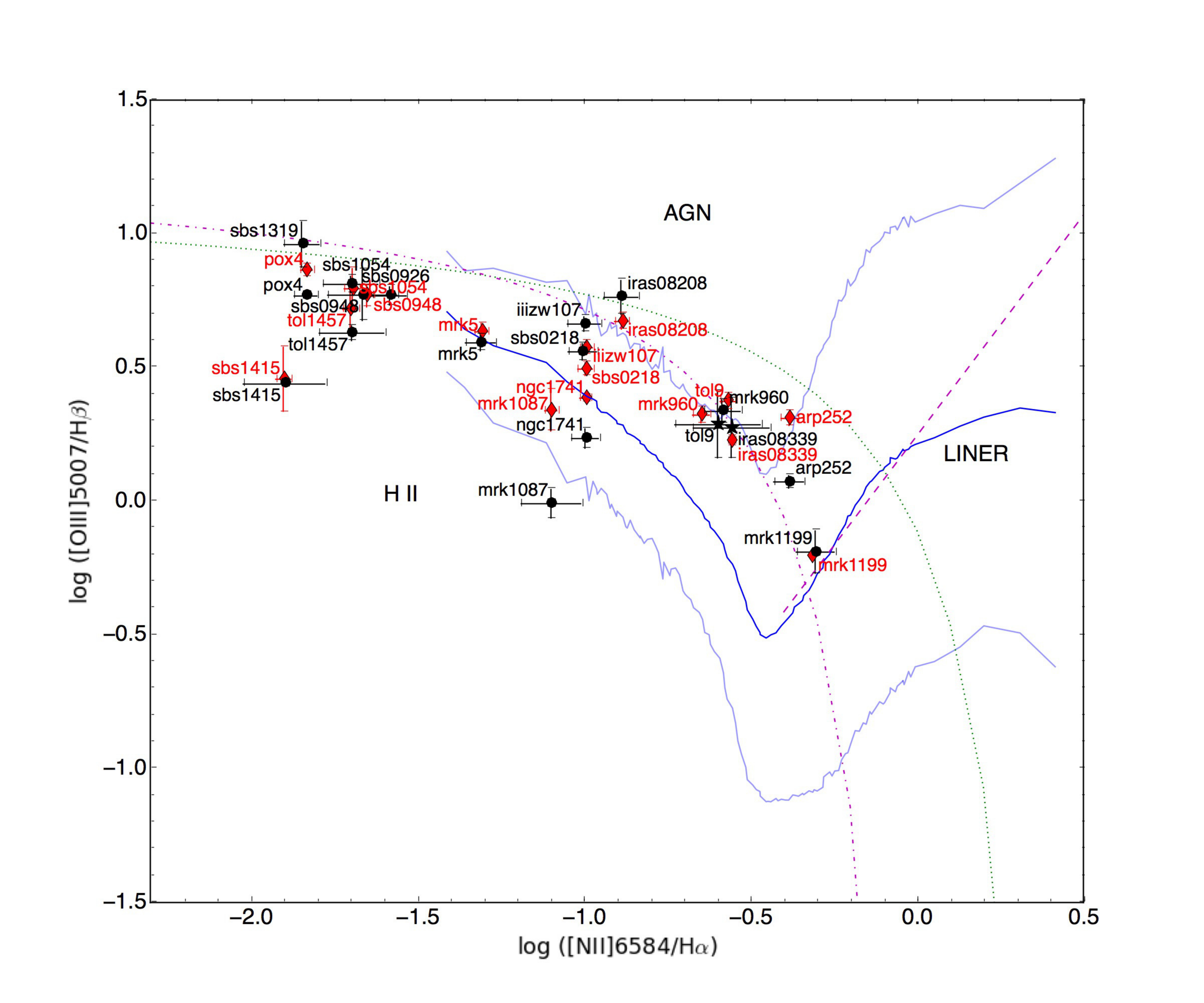}\\
\caption[fig_COcldy-eps-converted-to.pdf]{
Behavior of log([\ion{O}{3}]/H$\beta$) to log([\ion{N}{2}]/H$\alpha$) from observed W-R galaxy sample. Filled circles represent the objects for which we obtained a \te\ measurement, the black stars represent those objects for which we used \te\ from the literature, and the blue triangles represent reference points of objects studied in great detail: data for I Zw 18 was taken from \citet{leb13}, data for the Sun was taken from \citet{asp09}, data for 30 Dor was taken from \citet{pei03}, and data for Orion was taken from \citet{est09}. The red diamonds represent the data from \citet{lop04a} for NGC 1741, \citet{lop04b} for Mrk 1087, \citet{lop06} for IRAS 08339$+$6517, and from \citet{lop09} for the rest of the objects. The blue solid curve is the SDSS outline for the median of log([\ion{O}{3}]/H$\beta$) to log([\ion{N}{2}]/H$\alpha$), while the light blue curves are the same $\pm 3 \sigma$, respectively. The green dotted line represents the theoretical upper star-formation limit from \citet{kew01}, the dash-dotted magenta line represents the lower limit for AGNs from \citet{kau03}, and the dashed magenta line represents the division between AGN and LINERS from \citet{kau03}.
\label{fbpt}}
\end{center}
\end{figure}

To corroborate that photoionization in our W-R galaxy sample is caused by massive stars, we created a [\ion{O}{3}]/H$\beta$ to [\ion{N}{2}]/H$\alpha$ diagram, commonly referred to as BPT \citep{bpt81} diagram. We were not able to separate the [\ion{N}{2}] lines from H$\alpha$, nonetheless, we used the [\ion{N}{2}] line intensities presented in \citet{lop04a, lop04b, lop06} and \citet{lop09}. It is important to note that the our observations seem to correspond to the center regions observed in the works of LSE, however the exact location of the slits most likely {changed}. Figure \ref{fbpt} shows the [\ion{O}{3}]/H$\beta$ to [\ion{N}{2}]/H$\alpha$ observed values for our sample, as well as the intensity ratios measured by L\'opez-S\'anchez et al. We also show the theoretical upper limit for starburst galaxies as given in \citet{kew01} as a dotted green line, the lower limit for active galactic nuclei (AGN) as presented in \citet{kau03} as a dash-dotted magenta line, and the division between AGN and low-ionization nuclear emission-line regions (LINERS) according to \citet{kau03} as a dashed magenta line. The bulk of the objects from our W-R galaxy sample fall on the star-formation region of the diagram, though there are two objects (SBS 1319$+$579 and IRAS 08208$+$2816) that lie in the mix region between the star-forming galaxies and AGNs \citep{ric16}. Mrk 1199 lies a bit farther from the \ion{H}{2} region loci than the rest of the sample. This is due to its low ionization degree (see Section \ref{corrDM} and values of oxygen ionization degree, OID, and excitation index, $P$, in Table \ref{corabs}); \citet{san15} present a more detailed study on this issue.

\begin{figure}
\begin{center}
\includegraphics[angle=0,scale=0.42]{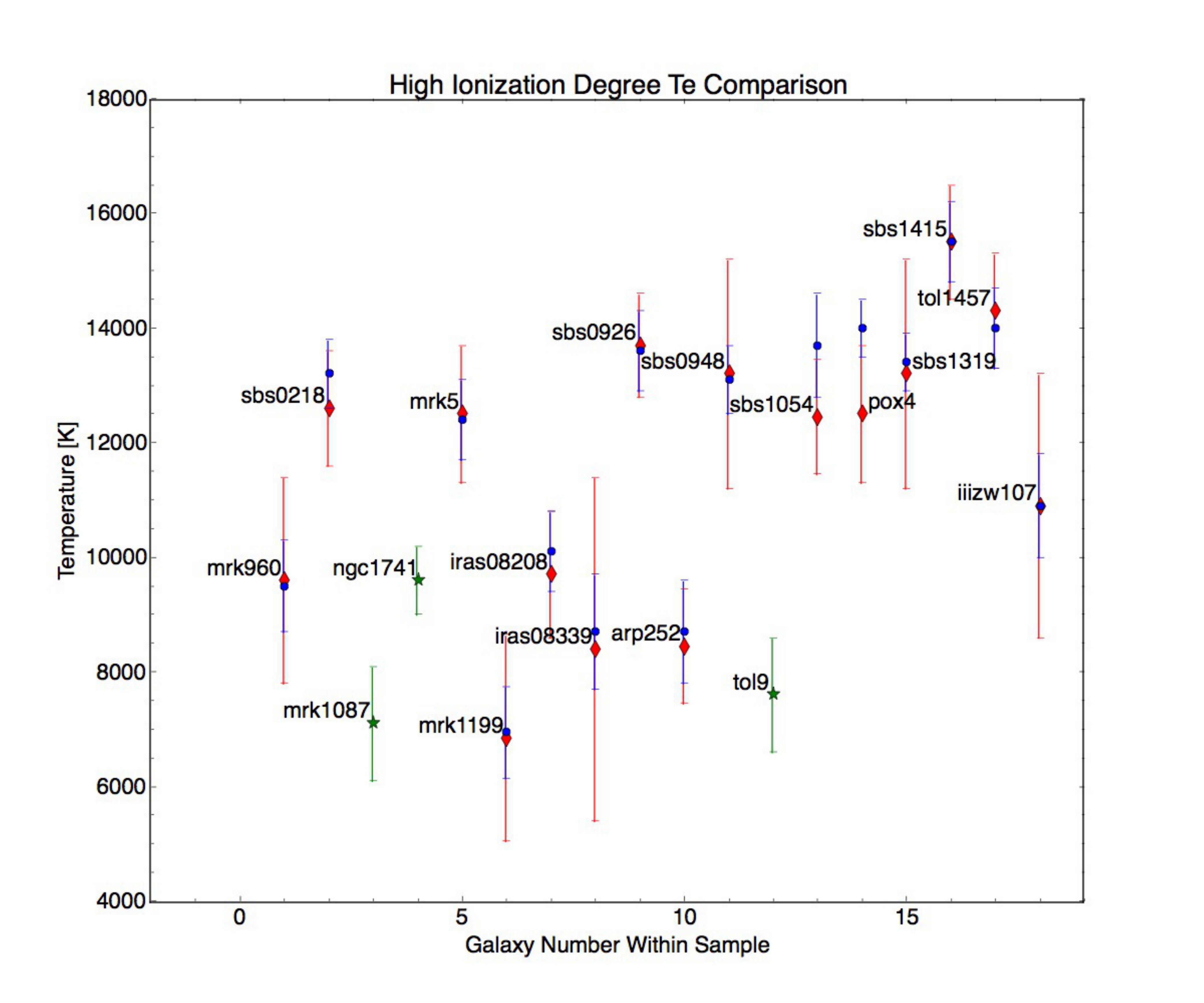}\\
\caption[fig_temp_comp-eps-converted-to.pdf]{
Comparison of our high-ionization zone temperatures versus those presented in LSE08, LSE09, LSE10a, LSE10b. Red diamonds represent our measurements, blue circles represent the data from LSE, and the green stars represent the temperatures from LSE we adopted.
\label{fTecomp}}
\end{center}
\end{figure}

We used the direct method to determine temperatures for our sample. For objects where [\ion{O}{3}] 4363 \AA\ or  [\ion{S}{3}] 6312 \AA\ were not observed or did not have good enough signal-to-noise (S/N), we used the temperature presented by \citet{lop09} or \citet{lop10b}. We carefully checked that the regions of the objects observed in our HST STIS data {were} the same as those observed in the works of L\'opez-S\'anchez \& Esteban; moreover, we find that our high ionization zone temperature determinations agree, within the errors, {with} those obtained by LSE08, LSE09, LSE10a, LSE10b. We were able to obtain a high ionization zone temperature measurement for 15 out of the 18 objects in our sample. Figure \ref{fTecomp} shows our high-ionization zone temperatures versus those obtained by LSE. 

The CELs of [\ion{O}{2}] 7320 and 7330 \AA\ and/or [\ion{S}{2}] 4069 and 4076 \AA\ were not observed or did not have good enough S/N (i.e. intensity uncertainty greater than 70\%) to determine temperatures for the low ionization zone (we defined a line with poor or low S/N as a line with an error greater than 40\%). To determine \te\ [\ion{O}{2}] we used the following relation taken from \citet{gar92}:
\begin{equation} 
T_e \textnormal{[\ion{O}{2}]} = 0.7 \times T_e \textnormal{[\ion{O}{3}]} + 3000.
\end{equation}
Garnett used the linear approximation to the relation between $T$(O$^+$) and $T$(O$^{+2}$) provided by \citet{cam86} from the models of \citet{sta82}, to determine ion-weighted mean electron temperatures.

It is important to keep in mind that the two-zone approximation is indeed a first order approximation to the actual thermal structure of the nebula (or \ion{H}{2} region). For the objects where [\ion{O}{3}] 4363 \AA\ or [\ion{S}{3}] 6312 \AA\ did not have good enough S/N (i.e. line had a width greater than 1.5 \AA\ and intensity uncertainty smaller than 50\%), we used both \te[\ion{O}{3}] and \te[\ion{O}{2}] as presented in \citet{lop10b}. The uncertainties in the temperatures and densities were obtained from PyNeb. 

To obtain electron density measurements we used the 6731/6711 [\ion{S}{2}] lines. In those objects where we could not determine one of the lines we adopted the electron density given in \citet{lop09} or \citet{lop10b}. We were able to obtain at least an upper electron density limit for 12 out of the 18 objects in our sample. The adopted electron temperatures and densities are presented in Table \ref{phys_props}.

\subsection{Chemical Abundances}\label{chemab}
In this section we describe how we obtained both the ionic and total gaseous abundances. We also briefly explain corrections to the direct method based on temperature inhomogeneities, dust depletion, and ionization structure.

\subsubsection{Ionic Abundances}
The ionic chemical abundances of He$^+$/H$^+$,  O$^{++}$/H$^+$, O$^+$/H$^+$, Ne$^{++}$/H$^+$, S$^{++}$/H$^+$, and S$^+$/H$^+$ were determined with the temperatures and densities shown in Table \ref{phys_props}. The resulting ionic abundances are given in Table \ref{ionchem_comp}. For the specific case of C$^{++}$/H$^+$, we followed the procedure described in \citep{gar95}.

\citet{gar95} used HST spectroscopy of dwarf galaxies to measure the relative abundances of C$^{+2}$/O$^{+2}$ from rest-UV emission lines \ion{C}{3}] 1909 \AA\ to \ion{O}{3}] 1666 \AA\ and \te[\ion{O}{3}]. The method assumes that the electron density of the \ion{H}{2} region in question is well below the critical densities for collisional de-excitation of both \ion{C}{3}] and \ion{O}{3}], $n_{crit}\sim$10$^5$ cm$^{-3}$ and $n_{crit}\sim$10$^3$ cm$^{-3}$, respectively \citep{ost05}. This is the case for all galaxies in our sample. The abundance of C$^{+2}$/O$^{+2}$ can then be computed in the low-density limit. The total abundance of carbon from the \citet{gar95} method depends on the temperature measured for the high ionization zone (see equation \ref{ec2o2}), hence the choice of the correct temperature is paramount for an accurate estimation of the total carbon abundance through this method.

The method described in \citet{gar95} is relatively straightforward once the collision strengths for C$^{+2}$ and O$^{+2}$ have been selected; we maintain the values adopted by \citet{gar95}. The method then essentially consists of four steps: (i) determine the ionic abundance ratio C$^{+2}$/O$^{+2}$ from the emission line ratio \ion{C}{3}] 1909 \AA\ to \ion{O}{3}] 1666 \AA\ and \te[\ion{O}{3}], (ii) determine the fraction of O$^{+2}$, $X$(O$^{+2}$), (iii) use Figure 2 in \citet{gar95} to obtain $X$(C$^{+2}$) and the ionization correction factor (ICF) for the unseen ions of carbon, ICF(C), and (iv) multiply the ionic abundance ratio C$^{+2}$/O$^{+2}$ by ICF(C) for obtaining C/O:
\begin{equation}\label{ec2o2}
\frac{\textnormal{C}^{+2}}{\textnormal{O}^{+2}}  =  0.089 \times e^{ \left ({\frac{-1.09}{T_e / 10^4}} \right )} \times \frac{I(\textnormal{\ion{C}{3}]}1909)}{I(\textnormal{\ion{O}{3}]} 1666)} 
\end{equation}
\begin{equation}
X\textnormal{(O}^{+2}\textnormal{)}  =  \frac{\textnormal{O}^{+2}}{ \textnormal{O}_{total}} 
\end{equation}
\begin{equation}
\textnormal{ICF(C)}   =  \left [ \frac{X\textnormal{(C}^{+2}\textnormal{)}}{X\textnormal{(O}^{+2}\textnormal{)}} \right ] ^{-1}
\end{equation}
\begin{equation}
\textnormal{C/O}  = \frac{\textnormal{C}^{+2}}{\textnormal{O}^{+2}} \times \textnormal{ICF(C)}.
\end{equation}
The constant 0.089 is the result of the product of the effective collision strengths between the two levels at electron temperatures below 20,000 K, $\Omega$(1661, 1666) \citep{bal81} and $\Omega$(1906, 1909) \citep{duf78}, the statistical weight of the corresponding lower level, the excitation potential of the transition, and the number density of the ion under consideration. For the objects where we did not obtain a measurement for \ion{O}{3}] 1666 \AA, we used the intensity of \ion{O}{3}] 1661 \AA, which is possible because the transitions of both lines arise from the same level as explained in \citet{gar95}.

\subsubsection{Total Abundances}
The total gaseous abundances for O, Ne, and S were determined with the following equations and the ICFs given in Table \ref{ICF}:
\begin{equation}   \frac{N({\rm O})}{N({\rm H})}=\frac{N({\rm O^{+}})+N({\rm O^{++}})}{N({\rm H^+})},
\label{totO}
\end{equation}
\begin{equation}
\frac{N({\rm Ne})}{N({\rm H})}=\frac{N({\rm O^{+}})+N({\rm O^{++}})}{N({\rm O^{++}})}\times \frac{N({\rm Ne^{++}})}{N({\rm H^+})} = {\rm ICF(Ne)} \times \frac{N({\rm Ne^{++}})}{N({\rm H^+})},
\label{totNe}
\end{equation}
and
\begin{equation}   \frac{N({\rm S})}{N({\rm H})}={\rm ICF(S)}\frac{N({\rm S^{+}})+N({\rm S^{++}})}{N({\rm H^+})}.
\label{totS}
\end{equation}

The resulting C abundances using the Garnett method, as well as abundances of for O, Ne, and S determined with the traditional analysis are presented in Table \ref{chem_comp}. We refer as traditional analysis to the standard assumption of a two-zone approximation to define the temperature structure of the object (with the electron temperature of [\ion{O}{3}] representing the high ionization potential ions, and that of [\ion{O}{2}] representing the low ionization potential ions) and the use of such temperatures to determine ionic and total abundances. Since the low resolution of the spectra does not permit to deblend the [\ion{N}{2}] lines from H$\alpha$, and the observed regions of the W-R galaxies of our sample match the center regions studied in the works of L\'opez-S\'anchez et al., we adopted the N abundances of \citet{lop08} and \citet{lop10b} and included these values in Table \ref{chem_comp} for completeness.  {We used standard error propagation equations to determine the final uncertainties from ours and those given in the L\'opez-S\'anchez et al. papers; nonetheless, additional sources of error may have been introduced.}

\subsubsection{Corrections to Direct Method}\label{corrDM}
The direct method has two essential shortcomings: (i) it depends on the capability to observe the weak auroral lines such as [\ion{O}{3}] 4363 \AA\ and [\ion{S}{3}] 6312 \AA, which can prove quite difficult in distant objects, with high redshift, or objects that are intrinsically faint; and (ii) the temperature structure of the object is not taken into account, i.e. abundances can be significantly underestimated with the direct method due to large and small scale temperature inhomogeneities \citep{pei67, pei69}. There are several works in the literature that address this problem, for a review see \citet{pei11} and \citet{lop12}. 

In this work we use the corrections proposed by \citet{pen12}. Pe\~na-Guerrero et al. used a sample of 28 \ion{H}{2} regions from the literature with measured temperature inhomogeneity parameter, $t^2$ \citep{pei67}, to derive a first approximation to the correction function of the O abundance determined with the auroral line [\ion{O}{3}] 4363 \AA, due to the thermal structure of the object as well as the fraction of oxygen depleted into dust grains; the authors refer to this relation as the Corrected Auroral Line Method (CALM): 12+log(O/H) \textsubscript{CALM} $= 1.0825 \times$(O/H)\textsubscript{Direct Method}$ - 0.375$. The authors then applied this correction function to the relations given by \citet{pil05} for upper and lower branches of the 12+log(O/H) vs R$_{23}$, where the strong line metallicity indicator $R_{23}=I$([\ion{O}{2}] 3727)$+I$([\ion{O}{3}] $4959+5007) / I$(H$\beta$), obtaining a strong line method that accounts for the thermal structure, dust depletion, and the ionization structure of the object. Pe\~na-Guerrero et al. refer to this technique as the Recalibrated $R_{23}$ Method (RRM). When the $R_{23}$ abundance falls in the between 8.29 and 8.55 (often called degeneracy zone), it is undefined which set of equations and values to use. In such cases, to know which set of equations is appropriate one can either use another metallicity indicator such as [\ion{N}{2}]/H$\alpha$ \citep{sto94} that increases linearly with 12+log(O/H), or use an indicator of the hardness of the ionizing radiation. \citet{pil00} proposed the excitation index, $P=I$([\ion{O}{3}] $4959+5007) /$($I$([\ion{O}{2}] 3727)$+I$([\ion{O}{3}] $4959+5007)$ as one such indicator. \citet{pen12} introduced the Oxygen Ionization Degree, OID=O$^{++}$/(O$^{+}+$O$^{++}$), which is a quantity equivalent to $P$. 

We used the the CALM and RRM methods to calculate the oxygen abundances corrected for dust depletion presented in Table \ref{corabs}. Column 1 is the galaxy name, columns 2 and 3 are  respectively, the carbon abundances determined through the method described in \citet{gar95} and corrected for dust depletion. Columns 4 through 7 are respectively, the oxygen abundances determined through: the direct method, dust depletion corrected as described in \citet{pei10}, CALM as described in \citet{pen12}, and RRM as described in \citet{pen12}. Column 8 is the strong line metallicity indicator $R_{23}$ as defined by \citet{pag79}. Column 9 shows the oxygen excitation ratio, $P$, as defined by \citet{pil00}, and column 10 presents the OID, as defined by \citet{pen12}.

\section{Discussion}\label{disc}
\subsection{C/O vs. O/H Diagram}\label{COdiagram}
The resulting C/O ratios and the gaseous oxygen abundance measured from our STIS observations are plotted in Figure \ref{fCOOH}. It is interesting to note that C/O in this figure and Figure 14 in \citet{hen06}, as well as N/O in Figure 12 of \citet{nav06} show an increase with respect to O/H starting around 12+log(O/H)$\sim$8.2. This behavior is likely due to the contribution to C and N by intermediate mass stars, which in turn implies that both carbon and nitrogen {could} be mainly produced in the same stars; {however, it is important to note that the slopes of the previously mentioned figures are quite different}.

The trend in Figure \ref{fCOOH} resembles that of the equivalent figure in \citet{gar95}, which shows an apparent increase in C/O with increasing O/H. Garnett et al. found that a good fit to their data was a power law of the form log(C/O)$=A+B$ log(O/H), with $A=1.01\pm0.39$ and $B=0.43\pm0.09$ for the abundance range $7.3\le\,$12+log(O/H)$\,\le8.7$. In this work we find that a power law may not be the best fit for our data. Though there is also an apparent increase of C/O with respect to O/H in our data, the behavior does not follow a specific curve, particularly when taking into account I Zw 18 \citep{leb13}, included as reference point along with 30 Doradus \citep{pei03}, Orion \citep{est09}, and the Sun \citep{asp09}. Nonetheless, we consider there is a section of the diagram (12+log(O/H)$>$7.5 or [O/H]$>$1.23) whose behavior could be described by the linear function
\begin{equation}\label{eqcooh}
\textnormal{log(C/O)} =  m\, \textnormal{log(O/H)} + b,
\end{equation}
where $m=1.8\pm0.4$ and $b=-14.9\pm2.9$, and the correlation coefficient is 0.78. The quantity log(O/H) is given in units of 12+log(O/H). {In this figure we also present the linear fit of \citet{gar95} and the data from \citet{ber16}, as well as their literature points, for comparison. The data from the low-metallicity high-ionization \ion{H}{2} regions in the Berg et al. work, as well as their literature data seem to agree better with the fit of Garnett. This could suggest that there is a dependence of the slope of the C/O versus O/H with respect to the IMF, since our sample has objects with a top-heavy IMF while as the Berg et al. sample does not. This, in turn, could imply that massive stars contribute more efficiently to the production of C in objects with a top-heavy IMF and with metallicity 12+log(O/H)$\gtrsim$8.0. }

Assuming a simple chemical evolution model with instantaneous recycling, the expected outcome for this plot would be a constant value for C/O. Such constant behavior would imply that either both C and O are primary elements, or that O is primary and C secondary but with C/O $\propto$ O/H. If we consider that only the primary carbon ``pollutes'' the ISM, an increase of C/O with increasing O/H would imply one or both of the following: (i) the instantaneous recycling approximation does not hold for both C and O, and (ii) the yield of C varies with metallicity.

\subsection{Behavior of C/N}
There does not seem to be a simple correlation between log(C/N) and 12+log(O/H) in the data if we consider the presence of I Zw 18, (Figure \ref{fCNOH}). Nonetheless, {omitting I Zw18}, there is a part of the diagram that {could} be described by a linear fit. We perform a linear fit to our data to describe such part of the diagram, we obtain 
\begin{equation}\
\textnormal{log(C/N)} =  m\, \textnormal{log(O/H)} + b,
\end{equation}
where $m=0.8\pm0.3$ and $b=-6.0\pm2.6$, and the correlation coefficient is 0.60. The quantity log(O/H) is given in units of 12+log(O/H). {If true, this linear increase of C/N with increasing O/H would imply that there is an additional contribution of C at higher metallicities, and an additional contribution of N at lower metallicities. This figure contrasts with Figure 6b of \citet{ber16}, in which the authors find a relatively constant behavior. The difference in our findings versus those in Berg et al. could be due to the physical differences of the samples used. They used low-metallicity and high-ionization \ion{H}{2} regions in dwarf galaxies while we have a range of both low and high ionization degrees and metallicities. Furthermore, our sample is composed of top-heavy IMF objects. Hence, the difference in the figures could suggest that the production of carbon and nitrogen in objects with such an IMF has a strong contribution from massive stars, and that these stars favor the production of C over N at metallicities higher than 12+log(O/H)$\gtrsim$8.0. Figure \ref{fCNOH} shows significant scatter, and the linear behavior seems to be only true for objects in our sample, again indicating that the origin of the C and N production is not homogeneous for objects with different IMFs.}

The C/N to O/H figure in \citet{gar95} appears to resemble a doubled-valued curve similar to a negative parabola, nonetheless they describe it as not showing a clear correlation. In our observations we did not have the resolution to separate the [\ion{N}{2}] emission lines from H$\alpha$, therefore, we adopted the nitrogen abundances derived by \citet{lop08} and \citet{lop10b}.  {We used standard error propagation equations to combine our uncertainties with those given in the L\'opez-S\'anchez \& Esteban papers. The reader should, however, be aware that additional sources of error may have been introduced. Figure \ref{fNCCH} shows a clear correlation for our STIS data, however, it is again evident that such behavior is not the same for objects with a different IMF.} We find that our data is consistent with a linear fit of the form: 
\begin{equation}\label{eqNCCH}
\textnormal{log(C/N)} = m\, \textnormal{log(C/H)} + b,
\end{equation}
with $b=-2.1\pm0.7$ and slope $m=0.4\pm0.1$, and a correlation coefficient of 0.68.

To determine if there is any relation between log(C/N) and log(N/O), we plotted these quantities in Figure \ref{fNOCN}. Though the correlation coefficient is very close to zero, we ran an MCMC model of a linear fit to our STIS data in order to {get a sense of the} possible fits. We used 100 walkers and did 500 runs. {As expected,} we found that about half of the fitted lines show a positive slope and half a negative slope. {As an experiment, we} took the average of the fits with negative slope to obtain a best estimate for parameters $m$ and $b$, and similarly we obtained another equation from the positive slopes. We then used these equations to determine carbon abundances for objects in the literature that have measurements of nitrogen and oxygen (see Section \ref{dla}). We find that only the C abundances determined from the equation with negative slope match the trend suggested by the {carbon} abundances determined from disk and halo stars (see Figure \ref{fCOOHext}). We plotted in Figure \ref{fNOCN} all the linear fits with negative slope (in red), as well as the {best fit determined by the MCMC algorithm} (in green). The equation for this line is
\begin{equation}\label{eqCNNO}
\textnormal{log(C/N)} = b - m\, \textnormal{log(N/O)},
\end{equation} 
where $b=0.31\pm^{0.15}_{0.12}$ and slope $m=0.21\pm^{0.09}_{0.11}$; {the uncertainties were determined from} the 25$^{th}$ and 75$^{th}$ percentiles. {Of course,} carbon abundances obtained with Equation \ref{eqCNNO} would be only a first {crude} approximation. {To determine the uncertainty of this equation, we calculated} the carbon abundances for our sample {and compared these with the abundances obtained with the Garnett method}. For the purpose of this analysis we will call benchmark values the C abundances obtained with the Garnett method. We compared the approximated carbon abundances with the benchmark values and we obtained an average difference of 0.38 dex. {Even though Equation \ref{eqCNNO} has a small statistical significance, it is interesting that the behavior of disk stars as well as the sample from \citet{ber16} also follow a negative slope. If this decrease of C/N with respect to N/O is true, further studies are needed to better characterize the behavior.}

\subsection{Abundance Discrepancy Factor}
Abundances of photoionized objects are generally determined using collisionally excited lines (CELs, see Section \ref{intro}). However, in bright objects, oxygen and carbon abundances can also be determined with recombination lines (RLs). A well-known problem in the chemical analysis of photoionized objects is the discrepancy between the abundances determined with RLs and those determined with CELs \citep[][and references therein]{pei93,gar07,pei07,est09,pen12,bla15}. This problem is generally referred to as the abundance discrepancy factor (ADF) problem, where ADF is defined as the ratio of abundances determined with RLs to those determined with CELs. 

RLs yield higher abundances than CELs. Typical ADF values for \ion{H}{2} regions lie in the 1.5 to 3 range \citep[e.g.][]{nic12, pei93, pei03, est09, pen12a, pen12} and in the 1.5 to 5 range (or higher than 20 in extreme cases) for most Planetary Nebulae \citep[e.g.][]{liu93, mcn13, nic12, pei14}. There have been two major explanations for {the ADF}: (i) high-metallicity inclusions that will create cool high-density regions surrounded by hot low-density regions \citep[e.g.][]{tsa05}, and (ii) thermal inhomogeneities in a chemically homogeneous medium that are caused by various physical processes such as shadowed regions, advancing ionization fronts, shock waves, magnetic reconnection, etc. \citep[e.g.][]{pei11}. A third explanation was recently proposed by \citet{nic12}: electrons depart from a Maxwell-Boltzmann equilibrium energy distribution but can be described with a ``$\kappa$-distribution''. Nicholls et al. suggest that a $\kappa \gtrsim 10$ is sufficient to encompass nearly all objects. 

Our STIS abundances do not have the necessary resolution to accurately obtain {abundances} for carbon and oxygen {via} RLs. \citet{est09} and \citet{est14} determined C and O abundances from RLs for a couple of objects in our sample. The comparison of the C and O abundances determined in this work with those determined by \citet{lop09} and \citet{est14} are presented in Table \ref{complit}, where column 1 is the galaxy name, column 2 the C abundances determined in this work with CELs, and column 3 the C abundances as determined in \citet{est14} RLs. Column 4 shows the oxygen abundances determined in this work from CELs, column 5 the O abundance as determined in \citet{lop09} also with CELs, and columns 6 and 7 present the O abundances as determined in \citet{est14} with CELs and RLs, respectively.

\subsection{Dust Depletion}\label{dust}
Depletion of heavy elements onto dust grains is important for the determination of accurate elemental abundances in the ISM \citep[e.g.][]{gar95, dwe98, est98, pei10, pen12}. In the case of oxygen, depletion has been shown to be dependent on metallicity \citep{pei10}: $0.09\pm0.03$ dex for 7.3$<$12+log(O/H)$<$7.8, $0.10\pm0.03$ dex for 7.8$<$12+log(O/H)$<$8.3, and $0.11\pm0.03$ for 8.3$<$12+log(O/H)$<$8.8. We have adopted this depletion correction for oxygen. However, the UV nature of the brightest carbon emission lines make a dust depletion study particularly difficult. Studies of C depletion suggest a correction for the nebular abundances from less than 0.1 \citep{sof94} to about 0.4 dex \citep{car93}. \citet{cun94} and \citet{gar95} recommend a correction of 0.2 dex, independent of metallicity. 

The dust corrected carbon and oxygen abundances {for our STIS sample} are shown in Table \ref{corabs}. The correction due to dust depletion is almost about twice as high for carbon than for oxygen. We decided not to plot the corrected abundances since {there is a possibility for both depletion corrections to be} dependent on metallicity. {If this is the case,} the correction on oxygen would be more accurate than that for carbon. Nonetheless, the overall shape of observed in Figure \ref{fCOOH} is preserved when using corrected abundances, though values are slightly increased. The behavior of all other figures also follows this description: we find no significant change in the overall shape of the curves presented in this work. {A possible consequence of the behavior of C/O versus O/H not to be flat could be that} depletion of carbon {has a} metallicity dependence. The contribution of carbon from stars more massive than 25M\solar\ to the ISM strongly depends on metallicity: the higher the mass the higher the C is expelled into the ISM \citet{mea92}. However, the higher the metallicity of gas, the greater the cross section for dust radiation in the UV \citep{gus99}, allowing for efficient destruction by photoionization. Hence, if there is a metallicity dependence in the C depletion, it is not a trivial one.

\subsection{Damped Lyman-$\alpha$ systems}\label{dla}
Damped Lyman-$\alpha$ systems (DLAs) are objects with high column density (log [$N$(\ion{H}{1}]) $\geq$ 20.3 cm$^{-2}$) of predominantly neutral gas detected in the spectra of an unrelated background light source, typically a quasar \citep{coo15}. DLAs have acquired particular attention mainly because: (i) the most metal-poor DLAs offer the unique opportunity to study the enrichment of galaxies due to the first generations of stars \citep{kob11}, and (ii) DLAs appear to sample various types of galaxies, from those with an extended \ion{H}{1} disk to subgalactic size halos \citep{wol05} at a wide range of redshifts.
 
The dominant neutral gas component of DLAs allows for the measurement of heavy element abundances to be straightforward, without the need for large ionization corrections. However, chemical abundances of carbon, nitrogen, and oxygen have received little attention in comparison to other heavy elements such as Cr, Fe, Mg, or Zn, though the relative abundances of C, N, and O particularly at low metallicities, provide extremely valuable information about early nucleosynthesis stages \citep[][and references therein]{wol05, coo15}. Since C and O are abundant elements with strong atomic transitions, their corresponding absorption lines are strongly saturated, thus making them unusable for abundance determination \citep{pet08}. The N absorption lines tend to be weak and blended with intergalactic Lyman-$\alpha$ (Ly$\alpha$) forest lines \citep{pet95, pet02}. Nonetheless, low metallicity and simple velocity structure DLAs facilitate the measurement of C, N, and O abundances \citep{pet08}. 

Several previous studies \citep[e.g.][]{des03, per07, pet08, coo11, coo15} have obtained C/O measurements from unsaturated \ion{C}{2} and \ion{O}{1} absorption lines. Furthermore, it has recently been suggested that DLAs have chemical evolution and kinematic structure that resembles that of Local Group dwarf galaxies \citep{coo15}. {We have included all the previously cited carbon measurements in the [C/O] versus [O/H] diagram, Figure \ref{fCOOHext}. In addition, we also included in this figure the sample of \citet{jam15}, which is a subsample of 12 extremely metal-poor galaxies morphologically selected from the SDSS, as well as the sample from \citet{nav06}, which is a compiled sample of low-metallicity emission-line galaxies. We calculated a crude first approximation to the C/O values for these two literature samples from their N/O and O/H values and Equation \ref{eqCNNO}. Even though the scatter is large in Figure  \ref{fCOOHext}, the overall shape of the figure with the resulting carbon abundances from Equation \ref{eqCNNO} seem to agree with similar figures in the literature, e.g. Figure 7 of \citet{ber16}.} Though uncertainties are large, if we take the center values to be true, it becomes apparent that different types of objects ``prefer" certain areas of the [C/O] vs. [O/H] diagram. For the carbon abundance we have gathered from the literature, and the ones we have derived in this work (either with the Garnett method or with our linear approximation), we observe that the most metal-poor DLAs are in the region $-0.7<$[C/O]$<0.7$ and [O/H] $<-1.8$, higher metallicity DLAs  are in the region $-1.5<$[C/O]$<0.0$ and $-2.5<$[O/H] $<-0.5$, extremely-low metallicity galaxies are in the region $-1.0<$[C/O]$<0.0$ and $-2.0<$[O/H] $<-0.5$, low-metallicity emission-line galaxies are in the region $-0.5<$[C/O]$<0.0$ and $-1.5<$[O/H] $<-0.5$, neutral ISM measurements with the 0.5 dex addition to make values comparable to those from star-forming regions \citep{jam14} are in the region $-2.0<$[C/O]$<-1.0$ and $-2.0<$[O/H] $<-0.5$, halo stars are in the region $-1.0<$[C/O]$<0.0$ and $-2.5<$[O/H] $<-0.5$, disk stars are in the region $-0.4<$[C/O]$<0.4$ and $-0.7<$[O/H] $<0.3$, and W-R galaxies are in the region $-1.5<$[C/O]$<1.0$ and $-1.3<$[O/H] $<0.0$. 
The translation of Equation \ref{eqcooh} into solar values is the following
\begin{equation}\label{eqcooh_solar}
\textnormal{[C/O]} = m\, \textnormal{[O/H]} + b,
\end{equation}
where $m=1.75\pm0.35$ and $b=0.69\pm0.22$. We have also translated the linear fit of \citet{gar95} to solar values, and we present it as well in Figure \ref{fCOOHext}. Metal abundances in the neutral ISM can be determined with far UV absorption lines, which requires bright UV sources to use as background spectra whose light is absorbed along the line of sight, similar to the study of DLAs \citep{lu96}. To compare our results with C determinations from absorption lines, it is important to consider that the analysis of the Far Ultraviolet Spectroscopic Explorer (FUSE) spectra of I Zw 18 with this technique indicates that the abundances of the alpha elements such as O, Ar, Si, and N are $\sim$0.5 dex lower in the neutral ISM than in the \ion{H}{2} regions, while the abundance of Fe remains the same \citep{alo03}, which has been confirmed by several other studies \citep[e.g.][and references therein]{leb09}.

Previous and current studies have reported large carbon enhancements in DLAs with C/O values matching that of halo stars of similar metallicity or even higher values, which is not expected from Galactic chemical evolution models based on conventional stellar yields \citep[e.g.][and references therein]{pet08, coo10, est14, coo15}. Such carbon enhancements suggest higher stellar carbon yields probably originated in stellar rotation, which promotes mixing in the stellar interiors \citep{pet08}. This could also be taken as independent confirmation of the non-flat behavior of C/O with respect to O/H as explained by metallicity-dependent stellar yields \citep{gar95}. Moreover, \citet{ake04} suggested that [C/O] values could not remain constant at [C/O]$=-0.5$, as previously thought, but increase again to approach solar metallicities at about [O/H]$\sim -3$, which would be due to metallicity-dependent non-LTE corrections to the [C/O] ratio. They proposed Population III stars as a possible explanation for the near-solar values of [C/O] at low metallicities, particularly if assuming a top-heavy IMF for these stars. Akerman et al. suggested that the higher temperatures reached in the cores of metal-free stars could shift the balance in the carbon and oxygen reactions, consequently producing a higher carbon yield, or that the mixing and fallback models of high energy supernova explosions of \citet{ume02, ume05} could be responsible for the carbon enhancement at early times. In either case, Figure \ref{fCOOHext} confirms the behavior predicted by \citet{ake04} for their ``standard" model, which uses the yields of \citet{mey02} for massive stars and those of \citet{van97} for low and intermediate mass stars, combined with the metal-free yields of \citet{chi02}. Furthermore, Figure \ref{fCOOHext} also supports the behavior of [C/O] versus [O/H] observed by \citet{ben06}, \citet{pet08}, \citet{est14}, {and \citet{ber16}}: \citet{ben06} suggest that the higher values of C/O at higher metallicities could be due in the most part by low and intermediate mass stars; \citet{pet08} pointed that the increase of the C/O ratio at lower metallicities suggest an additional source of carbon from the massive stars responsible for early nucleosynthesis; \citet{est14} explain that the position of star-forming dwarf galaxies coincides with that of Galactic halo stars suggests the same origin for the bulk of carbon in those galaxies, {and \citet{ber16}, argue that variations in the IMF could contribute to the large dispersion in the C/O values.}

The characteristic \ion{H}{1} Ly$\alpha$ absorption lines observed in DLAs are broadened by radiation damping, yet in some objects emission lines can be observed too. Of these emission lines, Ly$\alpha$ (1216 \AA) is the most valuable spectroscopic star-forming indicator in the redshift range of $4<z<6$ \citep{sta11}. At higher redshifts this line is not a reliable star-forming indicator anymore due to the resonant scattering by neutral gas in the IGM \citep{zit15}. The galaxy population at $z>6$ has in general lower UV luminosities and stellar masses than those from samples at $z\simeq2-3$, as well as large star formation rates, indicating a rapidly growing young stellar population \citep[e.g.][and references therein]{sta15}. Among the strongest emission lines of early galaxies are [\ion{O}{3}] 5007 \AA\ and H$\alpha$, however at $z\sim6$ these lines are situated at about $3-5 \mu$m, which makes them non-detectable with ground-based telescopes. Nonetheless, other UV emission lines such as \ion{O}{3}] 1660+6 and \ion{C}{3}] 1907+09 can probe the ionizing spectrum of galaxies at $z\gtrsim7$. \citet{sta14b} reported tentative detections of \ion{C}{3}] in two galaxies with $z$ of 6.029 and 7.213 (from Ly$\alpha$), while \citet{zit15} reached a 5$\sigma$ median flux limit for \ion{C}{3}] for an integration of 5 hours in the H-band in their pilot survey of the reionization era. This suggests that in the near future, points with $z>6$ will be added to the C/O versus O/H diagram giving us a more extended view of the carbon enrichment of the Universe.

\subsection{MCMC Modeling}

{Due to the low resolution of our STIS observations, we wanted an independent way to determine if our measured carbon through the Garnett method and oxygen abundances through the direct method were sensible. We used the MCMC technique to explore the parameter space and see where our carbon and oxygen abundances lie with respect to several thousand independent modeled samples with similar physical conditions. }

{We used \cldy\ for the photoionization models and a ionization spectrum input from {\tt Starburst99}. The chain ran using the {\tt emcee} algorithm. We obtained about 30,000 photoionization models per object. We calculated $\chi^2$ from comparing the observed line intensities to the modeled ones. We used the intensity relative to H$\beta$ of 9 lines to determine a $\chi^2$ value per model: \ion{H}{1} 4340, 4861, and 6563, \ion{He}{1} 5876, \ion{He}{2} 4686, [\ion{O}{2}] 3727, [\ion{O}{3}] 5007, \ion{C}{3}] 1909, and [\ion{S}{3}] 9532 \AA. To account for the observed oxygen temperatures, we made a sub-sample of models taking only those with temperatures of [\ion{O}{3}] and [\ion{O}{2}] $T\textsubscript{model}\leq$T\textsubscript{observed}$\pm2500\;$K, and we took the average of this sub-sample. The uncertainties for the average final abundances were obtained from the $25^{th}$ and $75^{th}$ percentiles. For the modeling set up and specific code versions used please see Section \ref{mcmc} in the appendix.}

{The bulk of the final average abundances obtained from the models for carbon and oxygen agree with our measured abundances, within the measurements' errors. Due to using almost no constraints for the runs we obtained large values for the $25^{th}$ and $75^{th}$ percentiles. Nonetheless, we noticed that the abundances of nitrogen, neon, and sulphur are not in close agreement to the observations. This is due to the loose constraints we used for the models and, since our STIS sample is composed of objects with a top-heavy IMF, it is a potential hint that the IMF we used for the models may not be the most adequate one. This issue requires further study to determine if it is true.}

{We combined all models from all objects in the sample to create a stacked log(C/O) vs. 12+log(O/H) diagram, and a log(N/O) vs. log(C/N) diagram. We noted that the main value of the models is between $-0.9$ and $-0.8$, which is interestingly coincidental with [C/O] for metal-poor halo stars according to \citet{tom92} (or $-0.9$ and $-0.8$ adopting the protosolar values for C/O\solar$\,=-0.26\pm0.07$ and 12+log(O/H)\solar=$8.73\pm0.05$ from \citet{asp09}), and with the carbon abundances determined via RLs of \citet{est14}. If the IMF does indeed play an important role in the C/O vs. O/H diagram, then this result would imply that there is a specific behavior for objects that have a similar IMF, and that halo stars are well described by a Kroupa IMF like the one we used for our models.}

\section{Summary and Conclusions}\label{conc}
We obtained STIS spectra covering the spectral region from about 1600 to 10,000 \AA\ for 18 starburst galaxies selected from the sample of W-R galaxies discussed by \citet{lop08,lop09,lop10a,lop10b}. Our goal is to study the enhancement of carbon in the ISM due to massive stars. We obtained physical conditions and chemical abundances for these 18 objects through standard nebular analysis. To determine the carbon abundances we used the method described in \citet{gar95}. The main results of the present work are:
\begin{enumerate}
\item{We confirm previous results: there is an increase in C/O with respect to O/H, yet we do not find a simple correlation. The most likely explanation for the non-constant relation (predicted as constant at low O/H by instantaneous recycling models for both carbon and oxygen) is that the yield of C varies with respect to O. Furthermore, our results indicate that the nucleosynthesis of carbon and/or oxygen deviates from the closed-box model, at least when dealing with objects with a clear ``starprint'' of massive stars (i.e. W-R stars). {This behavior agrees with the results of \citet{lop10c}, who also found that their galaxy sample did not agree with the closed-box model. These authors argue that the pristine gas inflow or the enriched gas outflow played an important role in the chemical evolution of their sample galaxies. When comparing our STIS sample C/O measurements with other references in the literature, such as \citet{ber16}, we find that there is a steeper slope of C/O vs O/H for our data, suggesting that the top-heavy IMF might have an effect on the carbon production, i.e. when massive stars are numerous, there is an additional contribution of C into the ISM for objects with metallicities higher than 12+log(O/H)$\gtrsim$8.0.}}
\item{Our data suggest that N/C ratio increases with increasing carbon abundance. {This is contradicts the behavior of the sample presented in \citet{ber16}, again suggesting that the IMF has a strong influence in the carbon production: at metallicities higher than 12+log(O/H)$\gtrsim$8.0 massive stars contribute more to the production of C. Further data is required to characterize this correlation, if it indeed exists.}}
\item{We find a potential {empirical} correlation between log(C/N) with respect to log(N/O). This relation estimates the carbon abundance from measurements of oxygen and nitrogen abundances, but should only be taken as a first order aproximation. The average difference of the carbon abundances approximated for our sample with this equation with respect to the carbon abundances obtained with the Garnett method {for the same sample}, is 0.38 dex.}
\item{In this work we used {an MCMC method determine if our carbon and oxygen abundance measurements were sensible. This method permits to explore the parameter space. However, to obtain accurate results with this technique, detailed photoionization models are required, but they can provide an effective and efficient technique to study correlations and/or degeneracies between abundances within an object as shown by \citet{tre04}, \citet{per14}, and \citet{bla15}.}  }
\item{The average value of log(C/O) from all \cldy\ models is about $-0.8$, which coincides very well with the main value of log(C/O) for halo stars. {If the IMF indeed has a strong effect in the production of carbon, the behavior shown by the \cldy\ models indicates the IMF of the models ``promoted'' a greater number of intermediate-mass stars rather than massive stars; hence the nucleosynthesis of carbon and nitrogen is most likely due to the same stars. The coincidence of C/O values could be an indication that  halo stars are well described by a Kroupa IMF.}}
\item{The addition of DLAs, disk and halo stars, and neutral ISM to the [C/O] versus [O/H] diagram provides additional insight into the carbon enrichment of the Universe with respect to oxygen. Independent results from different types of objects may confirm that the observed trends are due to stellar yields being metallicity-dependent rather than the instantaneous recycling assumption not holding true.}
\item{{From} the carbon determinations we compiled from the literature and those we determined in this work, we observe that different type of objects {seem to be} located in specific regions of the [C/O] versus [O/H] diagram. This diagram confirms the suggested behavior of [C/O] at lower metallicities observed by \citet{pet08} and \citet{est14}, and predicted by \citet{ake04}, which is likely due to Population III stars, before nucleosynthesis from Population II takes over, {and agrees with \citet{ber16} that the scatter of the C/O values are likely due to differences in the IMFs.}}
\end{enumerate}

Our results indicate that carbon and/or oxygen nucleosynthesis deviates from the instantaneous recycling and closed box models, at least in the presence of a large number of massive stars. The difference in the steep slope we find in the behavior of log(C/O) with respect to 12+log(O/H) versus previous studies of C/O in other objects, suggest that the carbon production is indeed greatly affected by the presence of massive stars. The behavior of C/O with metallicity resembles the relation between a primary and a secondary element, where the abundance ratio of the secondary to the primary element is predicted to increase with the abundance of its seed. A classic example of such behavior is for the N/O ratio to O/H, e.g. \citet{nav06}. The most plausible explanation for this behavior between C/O and O/H is that carbon is returned to the ISM by intermediate-mass stars on longer time-scales compared to oxygen, which is mainly returned to the ISM by massive stars; hence, C/O increases as O/H increases. This effect is amplified by the metallicity dependence of the carbon yields. Nonetheless, our measurements indicate that intermediate mass stars play a dominant role in the production of carbon in the range of $-2.5\lesssim$[O/H]$\lesssim-0.5$.

\vspace{8mm}
\emph{ACKNOWLEDGEMENTS}.

We are grateful to an anonymous referee for a careful reading of the manuscript and several useful suggestions. Support for this work has been provided by NASA through grant number O-1551 from the Space Telescope Science Institute, which is operated by AURA, Inc., under NASA contract NAS5-26555. 

This research has made use of the NASA/IPAC Extragalactic Database (NED) which is operated by the Jet Propulsion Laboratory, California Institute of Technology, under contract with the National Aeronautics and Space Administration.

\clearpage

% TABLES
\begin{center}
\begin{deluxetable}{lccccccccccccccc}
\tabletypesize{\tiny}
%\rotate
\tablecaption{
Main Properties of the W-R Galaxy Sample Analyzed in this Study.\label{main_props} }
\tablewidth{0pt}
\tablehead{
\colhead{Galaxy Name} &
\colhead{R.A. (2000)} &
\colhead{Dec. (2000)} &
\colhead{$E_{(B-V)_G}$\tablenotemark{a}} &
\colhead{$d$ [Mpc]\tablenotemark{b}} &
\colhead{Other names}
}
\startdata
Mrk 960			& 00 48 35.4 & $-$12 42 60 & 0.021 & $85.9\pm6.0$ & Haro 15, LEDA 2845, IRAS 00460$-$1259\\
SBS 0218$+$003	& 02 20 54.5 & $+$00 33 23 & 0.032 & $235.6\pm16.5$ & UM 420 \\
Mrk 1087			& 04 49 44.5 & $+$03 20 03 & 0.056 & $112.1\pm7.8$ & II Zw 23, IRAS 04470$+$0314\\
NGC 1741		& 05 01 37.4 & $-$04 15 31 & 0.045 & $54.6\pm3.8$ & HCG 31 AC, Mrk 1089, SBS 0459-043 \\
Mrk 5			& 06 42 15.5 & $+$75 37 30 & 0.076 & 13.2\tablenotemark{c} & SBS 0720$+$756, LEDA 19459\\
Mrk 1199			& 07 23 43.5 & $+$33 26 31 & 0.047 & $57.7\pm4.0$ & SBS 0720$+$20335, LEDA 20911\\
IRAS 08208$+$2816	& 08 23 54.9 & $+$28 06 22 & 0.028 & $193.9\pm13.6$ & WR 238 \\
IRAS 08339$+$6517	& 08 38 23.1 & $+$65 07 15 & 0.083 & $81.5\pm8$ & LEDA 24283 \\
\enddata
\tablenotetext{a} {Galactic reddening values calculated from the NASA/IPAC Extragalactic Database (NED) with $R_V=3.1$.}
\tablenotetext{b} {Distance taken from NED with $H_0 = 73\pm5$ km s$^{-1}$ Mpc$^{-1}$ and the Virgo, GA, and Shapley model.}
\tablenotetext{c} {Distances taken from \citet{zha13}.}
\tablenotetext{d} {Distance taken from \citet{lop08}.}
\end{deluxetable} 
\end{center}
\clearpage

\begin{center}
\begin{deluxetable}{lccccccccccccccc}
\tabletypesize{\small}
%\rotate
\tablecaption{
HST STIS Observation Log of the W-R Galaxy Sample Analyzed in this Study\label{tobs} }
\tablewidth{0pt}
\tablehead{
\colhead{Galaxy Name} &
\colhead{Observation} &
\colhead{Position} &
\colhead{Orbits} &
\multicolumn{4}{c}{Exposure Time [s] per Setting} \\
\cline{5-8}
&Date&Angle\tablenotemark{a}&&& Gratings && Mirror\tablenotemark{b}\\
\cline{5-7}
&&[deg] &&  G230L & G430L & G750L  }
\startdata
Mrk 960			&2012-09 & -76   & 1 & 1270 & 300 & 150 & 20\\ 
SBS 0218$+$003	&2012-12 & 29    & 2 & 3915 & 500 & 250 & 20\\
Mrk 1087			&2012-01 & -3     & 1 & 1315 & 300  &150 & 20\\
NGC 1741		&2012-01 &10     & 1 & 1155 & 300 & 150 & 60\\
Mrk 5			&2012-11 & -180 & 1 & 1555 & 300 & 150 & 60\\
Mrk 1199			&2012-11 & -144 & 1 & 1420 & 300 &150 & 5\\
IRAS 08208$+$2816	&2012-01 & -162 & 1& 1300 & 300 & 150 & 30\\
IRAS 08339$+$6517	&2013-01 & 144  & 1 &  1720 & 300 & 150 & 10\\
SBS 0926$+$606A	&2013-02 & 117  &  2 & 4250 & 600 & 300 & 60\\
Arp 252			&2013-03 &  18   & 1 & 1370 & 300 & 150 & 10\\
SBS 0948$+$532	&2013-03 & 117  & 4 & 8550 & 1842 & 900 & 120\\
Tol 9				&2012-05 & 49    & 2 & 4085 & 500 & 250 & 10\\
SBS 1054$+$365	&2013-05 & 59    & 2 & 4090 & 500 & 250 & 30\\
POX 4			&2012-04 & 25    & 1 & 1170 & 300 & 150 & 60\\
SBS 1319$+$579	&2012-08 & 39    & 4 & 8745 & 1899 & 912 & 120\\
SBS 1415$+$437	&2014-01 & -133 & 2 & 4170 & 500 & 250 & 30\\
Tol 1457$-$262		&2012-06 & 15    & 1 & 1340 & 300 &150 & 20\\
III Zw 107			&2012-01 & 0      & 1 & 1340 & 300 & 150 & 20\\
%\hline
\enddata
\tablenotetext{a} {Position angles are with reference to the north (upper edge of acquisition images, with East -positive- towards left side).}
\tablenotetext{b} {Seconds to obtain the acquisition images.}
\end{deluxetable} 
\end{center}
\clearpage

\begin{deluxetable}{lcccccccccccccccccccccccccccc}
\tabletypesize{\scriptsize}
%\rotate
\tablewidth{0pt}
\tablecaption{Line Intensities Ratios for Mrk 960, SBS 0218$+$003, Mrk 1087, NGC 1741, Mrk 5, and Mrk 1199
\label{tlines1}}
\tablehead{
\colhead{$\lambda$ (\AA)} & \colhead{Id.} & \colhead{$f_{\lambda}$} &&\multicolumn{6}{c}{$I(\lambda)/I$(H$\beta$),  $I$(H$\beta$)=100} \\
\cline{5-10}\\
&&&& Mrk 960 & SBS 0218$+$003 & Mrk 1087 & NGC 1741 & Mrk 5 & Mrk 1199\\
}
\startdata
%                                                                         mrk960                    sbs0218 	          mrk1087              ngc1741             mrk5                  mrk1199    
1661     & \ion{O}{3}]                &  1.124 &&     \nd             &    4$\pm$2  &    \nd &              1:  &    \nd  &     \nd  \\
1666     & \ion{O}{3}]                &  1.122 &&     13$\pm$6  &              \nd  &    1:  &    \nd  &    4$\pm$2  &        1:      \\
1883     & \ion{Si}{3}]               &  1.187 &&      12$\pm$1    &       6: &      29$\pm$3    &       9$\pm$1    &       9$\pm$1    &              8:   \\
1892     & \ion{Si}{3}]               &  1.197 &&      16$\pm$2    &              \nd  &              \nd  &               1: &      15$\pm$1    &              \nd  \\
1902     & \ion{Fe}{2}                &  1.208 &&      14$\pm$1    &              \nd  &      22$\pm$2    &       8$\pm$1    &      17$\pm$1    &      13$\pm$4    \\
1909     & \ion{C}{3}]                &  1.214 &&     85$\pm$32  &     8$\pm$2  &     5$\pm$2 &    11$\pm$2  &    10$\pm$4  &5$\pm$1    \\
2322     & [\ion{O}{3}]               &  1.337 &&    \nd   &       6: &       3: &       3: &       5: &              3:   \\
2471     & [\ion{O}{2}]               &  0.999 &&       1: &       2: &       1: &       6$\pm$1    &       1: &              6:   \\
\enddata
\tablenotetext{a} {Units of $F$(H$\beta$) are given in erg s$^{-1}$ cm$^{-2}\times10^{-15}$.}
\tablenotetext{b} {Units of equivalent widths (EWs) are given in \AA.}
\end{deluxetable}
\clearpage

\begin{deluxetable}{lccccccccccccccc}
\tabletypesize{\scriptsize}
%\rotate
\tablewidth{0pt}
\tablecaption{Line Intensities Ratios for IRAS $08208+2816$, IRAS $08339+6517$, SBS 0926$+$606A, Arp 252, SBS $0948+532$, and Tol 9
\label{tlines2}}
\tablehead{
\colhead{$\lambda$ (\AA)} & \colhead{Id.} & \colhead{$f_{\lambda}$} &&\multicolumn{6}{c}{$I(\lambda)/I$(H$\beta$),  $I$(H$\beta$)=100} \\
\cline{5-10}\\
&&&& IRAS 08208$+$2816 & IRAS 08339$+$6517 &SBS 0926$+$606A & Arp 252 & SBS 0948$+$532 & Tol 9
}
\startdata
%                                                                   iras08208              iras08339 	         sbs0926                 arp252              sbs0948                      tol9                             
1661     & \ion{O}{3}]                &  1.124 &&              \nd  &              \nd  &              \nd  &              \nd  &     9$\pm$3  &              \nd  \\
1666     & \ion{O}{3}]                &  1.122 &&     1: &     1:  &     8$\pm$3  &  1: &     6$\pm$1  &     0.16:   \\
1883     & \ion{Si}{3}]               &  1.187 &&       6$\pm$1    &       7$\pm$3    &       7$\pm$1    &       2: &               5: &              4:   \\
1892     & \ion{Si}{3}]               &  1.197 &&               1: &              \nd  &               2: &       4: &       9$\pm$2    &              \nd  \\
1902     & \ion{Fe}{2}                &  1.208 &&       5$\pm$1    &               1: &      20$\pm$2    &       6: &      25$\pm$4    &       8$\pm$2    \\
1909     & \ion{C}{3}]                &  1.216 &&      6$\pm$2  &    9$\pm$3  &    32$\pm$10  &     5$\pm$1  &    28$\pm$3  &     7$\pm$2  \\
2322     & [\ion{O}{3}]               &  1.337 &&       2$\pm$1    &              \nd  &       4$\pm$1    &       2: &       8$\pm$2    &              2:   \\
2471     & [\ion{O}{2}]               &  0.999 &&       3: &              \nd  &       1: &       3: &               2: &      18$\pm$4    \\
\enddata
\tablenotetext{a} {Units of $F$(H$\beta$) are given in erg s$^{-1}$ cm$^{-2}\times10^{-15}$.}
\tablenotetext{b} {Units of equivalent widths (EWs) are given in \AA.}
\end{deluxetable}
\clearpage

\begin{deluxetable}{lccccccccccccccc}
\tabletypesize{\scriptsize}
%\rotate
\tablewidth{0pt}
\tablecaption{Line Intensities Ratios for SBS $1054+365$, POX 4, SBS 1319$+$579, SBS 1415$+$437, Tol 1457$-$262, and III Zw 107
\label{tlines3}}
\tablehead{
\colhead{$\lambda$ (\AA)} & \colhead{Id.} & \colhead{$f_{\lambda}$} &&\multicolumn{6}{c}{$I(\lambda)/I$(H$\beta$),  $I$(H$\beta$)=100} \\
\cline{5-10}\\
&&&& SBS 1054$+$365 & POX 4 & SBS 1319$+$579 &  SBS 1415$+$437 & Tol 1457$-$262 &III Zw 107
}
\startdata
%                                                                        sbs1054                POX4 	           sbs1319               sbs1415                 tol1457                iiizw107    
1661     & \ion{O}{3}]                &  1.124 &&     2$\pm$1  &         \nd    &        4$\pm$1  &    2:  &    \nd  &              \nd  \\
1666     & \ion{O}{3}]                &  1.122 &&     5$\pm$1  &             5: &     10$\pm$4   &     7$\pm$3  &    7$\pm$4 &            2$\pm$1   \\
1883     & \ion{Si}{3}]               &  1.187 &&      20$\pm$2    &      11$\pm$1    &       4$\pm$1    &      15$\pm$2    &       5: &      20$\pm$1    \\
1892     & \ion{Si}{3}]               &  1.197 &&      18$\pm$2    &       5$\pm$1    &       5$\pm$1    &              \nd  &       6: &              \nd  \\
1902     & \ion{Fe}{2}                &  1.208 &&      20$\pm$2    &       7$\pm$1    &      31$\pm$2    &       6$\pm$2    &       1: &       6:   \\
1909     & \ion{C}{3}]                &  1.216 &&    28$\pm$8  &    23$\pm$7  &    59$\pm$18  &    4$\pm$1&     6$\pm$1  &    13$\pm$4  \\
2322     & [\ion{O}{3}]               &  1.337 &&       3$\pm$1    &       7$\pm$1    &       5$\pm$1    &       4$\pm$1    &       2: &       8:   \\
2471     & [\ion{O}{2}]               &  0.999 &&              \nd  &       4$\pm$1    &       2: &       2$\pm$1    &       1: &       3:   \\
\enddata
\tablenotetext{a} {Units of $F$(H$\beta$) are given in erg s$^{-1}$ cm$^{-2}\times10^{-15}$.}
\tablenotetext{b} {Units of equivalent widths (EWs) are given in \AA.}
\end{deluxetable}
\clearpage

\begin{center}
\begin{deluxetable}{lccccccccccccccc}
\tabletypesize{\small}
%\rotate
\tablecaption{
Physical Properties of the Observed Regions in Our W-R Galaxy Sample\label{phys_props} }
\tablewidth{0pt}
\tablehead{
\colhead{Galaxy Name} &
\colhead{\te \textsubscript{ High ionization zone} [K]} &
\colhead{\te \textsubscript{ Low ionization zone} [K]} &
\colhead{Electron Density [cm$^{-3}$]} &
}
\startdata
Mrk 960		          & $9,600\pm1,800$   & $9,700\pm1,800$\tablenotemark{b}     & $<100$\\
SBS 0218$+$003	 & $12,600\pm1,000$  & $11,800\pm1,000$\tablenotemark{b} & $<600$\\
Mrk 1087      		 & $7,100\pm1,000$\tablenotemark{a}& $8,000\pm1,100$\tablenotemark{a}  & $120\pm90$\\
NGC 1741	 	 & $9,600\pm600$\tablenotemark{a} & $9,600\pm600$\tablenotemark{a}  & $<100$\\
Mrk 5		      	 & $12,500\pm1,200$ & $11,750\pm1,200$\tablenotemark{b} & $<200$\\
Mrk 1199      		 & $6,850\pm1,800$   & $7,800\pm1,800$\tablenotemark{b}    & $<400$\\
IRAS 08208$+$2816	 & $9,700\pm1,100$   & $9,800\pm1,000$\tablenotemark{b}    & $<300$\\
IRAS 08339$+$6517	 & $8,400\pm3,000$   & $8,900\pm3,050$\tablenotemark{b}    & $<100$\\
\enddata
\tablenotetext{a} {Adopted from values determined by \citet{lop09,lop10b}. }
\tablenotetext{b} {Values obtained using Equation 2 in \citet{gar92}.}
\end{deluxetable}
\end{center}
\clearpage

\begin{center}
\begin{deluxetable}{lcccccccccccccccccccccccc}
\tabletypesize{\scriptsize}
%\rotate
\tablecaption{
Ionic Chemical Composition of the Observed Regions in Our W-R Galaxy Sample\tablenotemark{*}  \label{ionchem_comp} }
\tablewidth{0pt}
\tablehead{
\colhead{Galaxy Name} &
 He$^+$/H$^+$  & O$^{++}$/H$^+$ & O$^+$/H$^+$ & C$^{++}$/H$^+$ & Ne$^{++}$&/H$^+$ S$^{++}$/H$^+$ & S$^+$/H$^+$
}
\startdata               
%                                     He              O3                      O2                        C3                   Ne3                         S3                         S2            
Mrk 960		          & 10.99: & 7.96$\pm$0.17& 7.59$\pm$0.22& 6.81$\pm$0.20& 7.43$\pm$0.20& 6.60$\pm$0.20& 6.53$\pm$0.20 \\
SBS 0218$+$003       & 10.92: & 7.80$\pm$0.06& 7.53$\pm$0.08& 6.84$\pm$0.19& 7.08$\pm$0.10& 5.95$\pm$0.05& 6.07$\pm$0.02\\
Mrk 1087		          & 10.88: & 8.07$\pm$0.14& 8.13$\pm$0.19& 8.58$\pm$0.19& 7.92$\pm$0.19&    \nd                & 6.67$\pm$0.22 \\  
NGC 1741                   & 10.97: & 7.86$\pm$0.09& 8.02$\pm$0.12& 7.83$\pm$0.09& 7.29$\pm$0.15& 6.39$\pm$0.10& 6.04$\pm$0.13 \\
Mrk 5		           & 10.87: & 7.84$\pm$0.09& 7.33$\pm$0.12& 6.95$\pm$0.19& 7.42$\pm$0.14& 6.22$\pm$0.09& 5.81$\pm$0.12  \\
Mrk 1199	                    & 10.83: & 8.04$\pm$0.19& 8.02$\pm$0.25& 8.89$\pm$0.23& 7.44$\pm$0.25& 6.52$\pm$0.22& 6.32$\pm$0.29 \\
IRAS 08208$+$2816	   & 11.01: & 8.37$\pm$0.10& 7.53$\pm$0.13& 7.54$\pm$0.15& 7.76$\pm$0.12& 6.39$\pm$0.11 & 5.85$\pm$0.15\\
IRAS 08339$+$6517   & 10.98: & 8.11$\pm$0.06& 8.05$\pm$0.09& 8.26$\pm$0.19& 8.02$\pm$0.09& 7.06$\pm$0.08& 5.76$\pm$0.06 \\
\enddata
\tablenotetext{*} {In units of 12 + log $n(X^{i+})/n$(H$^+$). Gaseous content only.}
\end{deluxetable}
\end{center}
\clearpage

\begin{center}
\begin{deluxetable}{lcccccccccccccccccccccccc}
%\tabletypesize{\small}
%\rotate
\tablecaption{
Ionization Correction Factors \label{ICF} }
\tablewidth{0pt}
\tablehead{
\colhead{Galaxy Name} 
 &&  C\tablenotemark{a} & Ne\tablenotemark{b} & S\tablenotemark{c}
}
\startdata
%                                                  C                        Ne                       S
Mrk 960		          &&  1.25$\pm$0.45 & 1.42$\pm$0.15 & 1.12$\pm$0.20 \\
SBS 0218$+$003       &&  1.02$\pm$0.45 & 1.53$\pm$0.13 & 1.12$\pm$0.20 \\
Mrk 1087		          &&  1.05$\pm$0.45 & 2.15$\pm$0.23 & 1.12$\pm$0.20 \\
NGC 1741                   &&  1.00$\pm$0.45 & 2.45$\pm$0.21 & 1.12$\pm$0.20 \\
Mrk 5		           &&  1.05$\pm$0.20 & 1.30$\pm$0.10 & 1.26$\pm$0.20 \\
Mrk 1199	                    &&  1.05$\pm$0.45 & 1.97$\pm$0.31 & 1.12$\pm$0.20 \\
IRAS 08208$+$2816	   &&  1.00$\pm$0.20 & 1.14$\pm$0.11 & 1.78$\pm$0.20 \\
IRAS 08339$+$6517   &&  1.05$\pm$0.45 & 1.86$\pm$0.40 & 1.12$\pm$0.20 \\ 
SBS 0926$+$606A	   &&  1.25$\pm$0.20 & 1.16$\pm$0.13 & 1.78$\pm$0.20 \\
Arp 252		           && 1.02$\pm$0.45 & 1.68$\pm$0.08 & 1.12$\pm$0.20 \\
SBS 0948$+$532        &&  1.11$\pm$0.20 & 1.18$\pm$0.39 & 1.26$\pm$0.20 \\
Tol 9			           &&  0.95$\pm$0.45 & 1.88$\pm$0.45 & 1.12 $\pm$0.20 \\ 
SBS 1054$+$365        &&  1.25$\pm$0.20 & 1.10$\pm$0.13 & 1.78$\pm$0.20 \\
POX 4		           &&  1.18$\pm$0.20 & 1.09$\pm$0.08 & 1.78$\pm$0.20  \\
SBS 1319$+$579	  &&  1.25$\pm$0.20 & 1.11$\pm$0.08 & 1.78$\pm$0.20 \\
SBS 1415$+$437	  &&  1.00$\pm$0.20 & 1.31$\pm$0.06 & 1.26$\pm$0.20  \\
Tol 1457$-$262	          &&  1.11$\pm$0.45 & 1.34$\pm$0.10 & 1.26$\pm$0.20 \\
III Zw107      	       	  &&  1.09$\pm$0.45 & 1.33$\pm$0.10 & 1.26$\pm$0.20  \\
%\hline
\enddata
\tablenotetext{a} {Obtained from \citet{gar95}.}
\tablenotetext{b} {Obtained from \citet{pei69}.}
\tablenotetext{c} {Obtained from \citet{gar89}.}
\end{deluxetable}
\end{center}
\clearpage

\begin{center}
\begin{deluxetable}{lcccccccccccccccccccccccc}
\tabletypesize{\scriptsize}
\rotate
\tablecaption{
Total Chemical Composition of the Observed Regions in Our W-R Galaxy Sample\tablenotemark{*}  \label{chem_comp} }
\tablewidth{0pt}
\tablehead{
\colhead{Galaxy Name} &
& O&	C&	N\tablenotemark{a}&	Ne& S & C/O & N/O & Ne/O & S/O 
}
\startdata
Mrk 960		           && 8.11$\pm$0.19& 7.49$\pm$0.20& 7.34$\pm$0.10& 7.59$\pm$0.20& 6.90$\pm$0.2 
                                                                  & -0.62$\pm$0.28& -1.39$\pm$0.28& -0.52$\pm$0.28& -1.21$\pm$0.28 \\
SBS 0218$+$003        && 7.99$\pm$0.06& 6.82$\pm$0.19& 6.84$\pm$0.06& 7.27$\pm$0.10& 6.36$\pm$0.09
                                                                  & -1.44$\pm$0.20& -1.15$\pm$0.08& -0.72$\pm$0.12& -1.63$\pm$0.11 \\
Mrk 1087		           && 8.41$\pm$0.16& 8.60$\pm$0.19& 7.76$\pm$0.16& 8.25$\pm$0.19& 6.79$\pm$0.19\tablenotemark{a} 
                                                                  & 0.19$\pm$0.25& -0.65$\pm$0.23& -0.16$\pm$0.25& -1.62$\pm$0.25 \\
NGC 1741                   && 8.25$\pm$0.10& 7.83$\pm$0.20& 7.10$\pm$0.09& 7.68$\pm$0.15& 6.60$\pm$0.11 
                                                                  & -0.42$\pm$0.22& -1.15$\pm$0.13& -0.57$\pm$0.18& -1.65$\pm$0.15 \\
Mrk 5		           && 7.98$\pm$0.10& 6.97$\pm$0.19& 6.72$\pm$0.09& 7.54$\pm$0.14& 6.46$\pm$0.10 
                                                                  & -0.99$\pm$0.21& -1.24$\pm$0.13& -0.42$\pm$0.17& -1.50$\pm$0.14 \\
Mrk 1199	                    && 8.33$\pm$0.22& 8.91$\pm$0.23& 7.90$\pm$0.18& 7.73$\pm$0.25&  6.78$\pm$0.25 
                                                                  & 0.58$\pm$0.32& -0.43$\pm$0.28& -0.70$\pm$0.33& -1.55$\pm$0.33 \\
IRAS 08208$+$2816	   && 8.43$\pm$0.11& 7.54$\pm$0.15& 7.44$\pm$0.10& 7.82$\pm$0.12& 6.75$\pm$0.13 
                                                                  & -0.89$\pm$0.19& -0.99$\pm$0.15& -0.61$\pm$0.16& -1.68$\pm$0.17 \\
IRAS 08339$+$6517   && 8.42$\pm$0.07& 8.27$\pm$0.19& 7.51$\pm$0.10& 8.25$\pm$0.09& 7.13$\pm$0.12 
                                                                  & -0.15$\pm$0.20& -0.91$\pm$0.07& -0.17$\pm$0.11& -1.29$\pm$0.14 \\
\enddata
\tablenotetext{*} {In units of 12 + log $n(X)/n$(H). Gaseous content only.}
\tablenotetext{a} {Values taken from the work of \citet{lop09, lop10b}. These abundances are presented for completeness, and to compare with modeled He abundances.}
\end{deluxetable}
\end{center}
\clearpage

\begin{center}
\begin{deluxetable}{lcccccccccccccccccccccccc}
\tabletypesize{\scriptsize}
%\rotate
\tablecaption{
Corrections to the Direct Method Carbon and Oxygen Abundances\tablenotemark{*}  \label{corabs} }
\tablewidth{0pt}
\tablehead{
\colhead{Galaxy Name} &
\multicolumn{2}{c}{12+log(C/H)} &&
\multicolumn{4}{c}{12+log(O/H)} &
\colhead{$R_{23}$} &
\colhead{$P$} &
\colhead{OID\tablenotemark{g}} \\
\cline{2-3}
\cline{5-8}
& This Work\tablenotemark{a} & +Dust\tablenotemark{b}  &&  
Direct\tablenotemark{c} & +Dust\tablenotemark{d} & CALM\tablenotemark{e} & RRM\tablenotemark{f}
}
\startdata
Mrk 960			&  7.49$\pm$0.20 & 7.69 && 8.11$\pm$0.19 & 8.21 & 8.41$\pm$0.21 & $dz$ & 3.89$\pm$0.19 & 0.75$\pm$0.08 & 0.70$\pm$0.10 \\
SBS 0218$+$003	&  6.82$\pm$0.19 & 7.02 && 7.99$\pm$0.06 & 8.09 & 8.27$\pm$0.06 & 8.10$\pm$0.12 & 6.71$\pm$0.32 & 0.72 $\pm$0.09& 0.65$\pm$0.10 \\
Mrk 1087			&  8.81$\pm$0.19 & 9.01 && 8.42$\pm$0.16 & 8.52 & 8.72$\pm$0.17 & 8.92$\pm$0.19 & 2.79$\pm$0.19 & 0.47$\pm$0.09 & 0.47$\pm$0.10 \\
NGC 1741		&  7.83$\pm$0.20 & 8.05 && 8.25 $\pm$0.10& 8.35 & 8.55$\pm$0.11 & 8.68$\pm$0.10 & 4.97$\pm$0.28 & 0.46$\pm$0.10 & 0.41$\pm$0.12 \\
Mrk 5			&  6.97$\pm$0.19 & 7.17 && 7.96$\pm$0.10 & 8.06 & 8.24$\pm$0.11 & 7.89$\pm$0.12 & 6.32$\pm$0.31 & 0.82$\pm$0.10 & 0.72$\pm$0.11 \\
Mrk 1199			&  8.91$\pm$0.23 & 9.11 && 8.33$\pm$0.22 & 8.44 & 8.64$\pm$0.24 & 9.05$\pm$0.24 & 1.79$\pm$0.22 & 0.48$\pm$0.09 & 0.51$\pm$0.10 \\
IRAS 08208$+$2816	&  7.54$\pm$0.15 & 7.74 && 8.43$\pm$0.11 & 8.54 & 8.75$\pm$0.12 & 8.66$\pm$0.12 & 8.61$\pm$0.51 & 0.90$\pm$0.11 & 0.87$\pm$0.12 \\
IRAS 08339$+$6517	& 8.27$\pm$0.19 & 8.47 && 8.42$\pm$0.07 & 8.53 & 8.74$\pm$0.08 & 8.90$\pm$0.09 & 3.99$\pm$0.40 & 0.63$\pm$0.16 & 0.67$\pm$0.19 \\
\enddata
\tablenotetext{*} {In units of 12+log $n(X)/n$(H). }
\tablenotetext{a} {Abundances determined with the method described in \citet{gar95}.}
\tablenotetext{b} {Consistent correction of 0.2 dex as discussed in the text, Section \ref{dust}. Uncertainties remain the same.}
\tablenotetext{c} {Determined from the Direct method, that is using $T_e$[OIII] as measured from the line ratio [\ion{O}{3}] 4363/5007.}
\tablenotetext{d} {Using metallicity dependent correction from 0.08 to 0.12 dex according to \citet{pei10}. Uncertainties remain the same as with the Direct Method.}
\tablenotetext{e} {Corrected Auroral Line Methodas described in \citet{pen12}.}
\tablenotetext{f} {Recalibrated $R_{23}$ Method as described in \citet{pen12}. An abundance in the degeneracy zone is marked as $dz$.}
\tablenotetext{g} {Oxygen Ionization Degree as defined in \citet{pen12}: OID = O$^{++}$ / (O$^{+} + $O$^{++}$).}
\end{deluxetable}
\end{center}
\clearpage

\begin{center}
\begin{deluxetable}{lcccccccccccccccccccccccc}
\tabletypesize{\small}
%\rotate
\tablecaption{
Comparison of Total C and O abundances with Literature\tablenotemark{*}.\label{complit} }
\tablewidth{0pt}
\tablehead{
\colhead{Galaxy Name} &
\multicolumn{2}{c}{12+log(C/H)} &&
\multicolumn{4}{c}{12+log(O/H)} \\
\cline{2-3}
\cline{5-8} 
& \colhead{This work\tablenotemark{a}} &  \colhead{Est14\tablenotemark{b}} 
&& \colhead{This work\tablenotemark{a}} & \colhead{Lop09\tablenotemark{b}} & \multicolumn{2}{c}{Est14\tablenotemark{b}} \\
\cline{7-8}
& CELs & RLs && CELs & CELs & CELs & RLs
}
\startdata
Mrk 960		             & 7.49$\pm$0.20 & -   && 8.11$\pm$0.19 & $8.37\pm0.10$ & - & - \\
SBS 0218$+$003	     & 6.82$\pm$0.19 & -  && 7.99$\pm$0.06 & $7.95\pm0.05$ & - & - \\
Mrk 1087			      & 8.60$\pm$0.19 & -   && 8.41$\pm$0.16 & $8.57\pm0.10$ & - & -  \\
NGC 1741		      & 7.83$\pm$0.20 & $<7.97$\tablenotemark{d} && 8.25$\pm$0.10 &  $8.22\pm0.05$ & $8.41\pm0.06$\tablenotemark{d} & $<8.68$\tablenotemark{d} \\
Mrk 5			      & 6.97$\pm$0.19 & -   && 7.96$\pm$0.10 & $8.07\pm0.07$ & - & - \\
Mrk 1199			      & 8.91$\pm$0.23 & -   && 8.33$\pm$0.22 & $8.57\pm0.21$ & - & - \\
IRAS 08208$+$2816	      & 7.54$\pm$0.15 & -   && 8.43$\pm$0.11 & $8.41\pm0.10$ & - & - \\
IRAS 08339$+$6517      & 8.27$\pm$0.19 & -   && 8.42$\pm$0.07 & $8.45\pm0.10$ & - & - \\
\enddata
\tablenotetext{*} {Comparison of the total C and O abundances derived in this work using the Direct Method with those previously reported in the literature. Abundances are presented in units of 12+log(X/H), and considering a homogeneous thermal structure ($t^2=0.000$).}
\tablenotetext{a} {Abundances determined with the method described in \citet{gar95}.}
\tablenotetext{b} {Abundances taken from \citet{est14}. The oxygen abundances presented from RLs are those values for O$^{2+}$.}
\tablenotetext{c} {Abundances taken from \citet{lop09}. }
\tablenotetext{d} {Abundances taken from \citet{est09} from the optical RL ionic abundance of X$^{2+}$. }
\tablenotetext{e} {Abundance taken from \citet{jam14} from the \ion{C}{2} UV absorption lines for the neutral ISM. The neutral ISM value is lower than that from  the \ion{H}{2} regions by $\sim$0.5 dex, which agrees with the results of \citet{alo03} and \citet{leb09}.}
\end{deluxetable}
\end{center}
\clearpage

\begin{center}
\begin{deluxetable}{lcccccccccccccccccccccccc}
\tabletypesize{\small}
%\rotate
\tablecaption{
Chemical Composition of the Modeled Sample\tablenotemark{*}  \label{modchem_comp} }
\tablewidth{0pt}
\tablehead{
\colhead{Galaxy Name} &
\colhead{He/H} & 
\colhead{O/H} & 
\colhead{C/H} & 
\colhead{N/H}& 
\colhead{Ne/H}& 
\colhead{S/H}
}
\startdata
Mrk 960		            & 10.89$\pm^{0.40}_{0.34}$& 8.50$\pm^{0.14}_{0.07}$& 8.29$\pm^{0.64}_{0.60}$& 7.34$\pm^{0.11}_{0.20}$& 8.04$\pm$0.21& 6.78$\pm^{0.20}_{0.17}$ \\
SBS 0218$+$003         &10.81$\pm$0.05& 7.86$\pm^{0.09}_{0.11}$& 6.90$\pm^{0.11}_{0.27}$& 6.79$\pm{0.14}$& 7.46$\pm$0.14& 6.09$\pm$0.18 \\
Mrk 1087			    & 10.65$\pm^{0.64} _{0.39}$& 8.46$\pm^{0.12} _{0.08}$& 8.84$\pm^{0.65} _{0.61}$& 7.37$\pm^{0.23} _{0.29}$& 7.97$\pm0.23$& 6.71$\pm$0.21 \\
NGC 1741		    & 10.99$\pm$0.48& 8.05$\pm$0.17& 7.60$\pm^{0.34}_{0.37}$& 6.99$\pm^{0.18}_{0.22}$& 7.58$\pm$0.20& 6.27$\pm$0.13 \\
Mrk 5			    & 10.70$\pm^{0.03}_{0.06}$& 8.09$\pm^{0.09}_{0.07}$& 7.62$\pm^{0.20}_{0.27}$& 7.10$\pm$0.14& 7.64$\pm^{0.12}_{0.09}$& 6.31$\pm$0.12 \\
Mrk 1199			    & 11.19$\pm^{0.52}_{0.63}$& 8.59$\pm^{0.08}_{0.05}$& 9.07$\pm^{0.38} _{0.27}$& 7.62$\pm^{0.14}_{0.28}$&8.48$\pm^{0.11} _{0.08}$& 7.03$\pm0.15$ \\
IRAS 08208$+$2816	    & 10.86$\pm$0.11& 8.51$\pm^{0.10}_{0.07}$& 7.51$\pm^{0.09}_{0.28}$& 7.66$\pm^{0.23}_{0.20}$& 8.19$\pm^{0.17}_{0.15}$& 6.69$\pm^{0.22}_{0.18}$ \\
IRAS 08339$+$6517    & 11.28$\pm^{0.37} _{0.29}$& 8.52$\pm^{0.23} _{0.10}$& 8.85$\pm^{0.60} _{0.54}$& 7.44$\pm$0.28& 8.23$\pm^{0.17} _{0.09}$& 6.79$\pm$0.16 \\
\enddata
\tablenotetext{*} {In units of 12+log $n(X)/n$(H).}
\end{deluxetable}
\end{center}
\clearpage

%FIGURES
\clearpage

\clearpage

\clearpage

\clearpage

\clearpage

\clearpage

\begin{figure}
\begin{center}
\includegraphics[angle=0,scale=0.22]{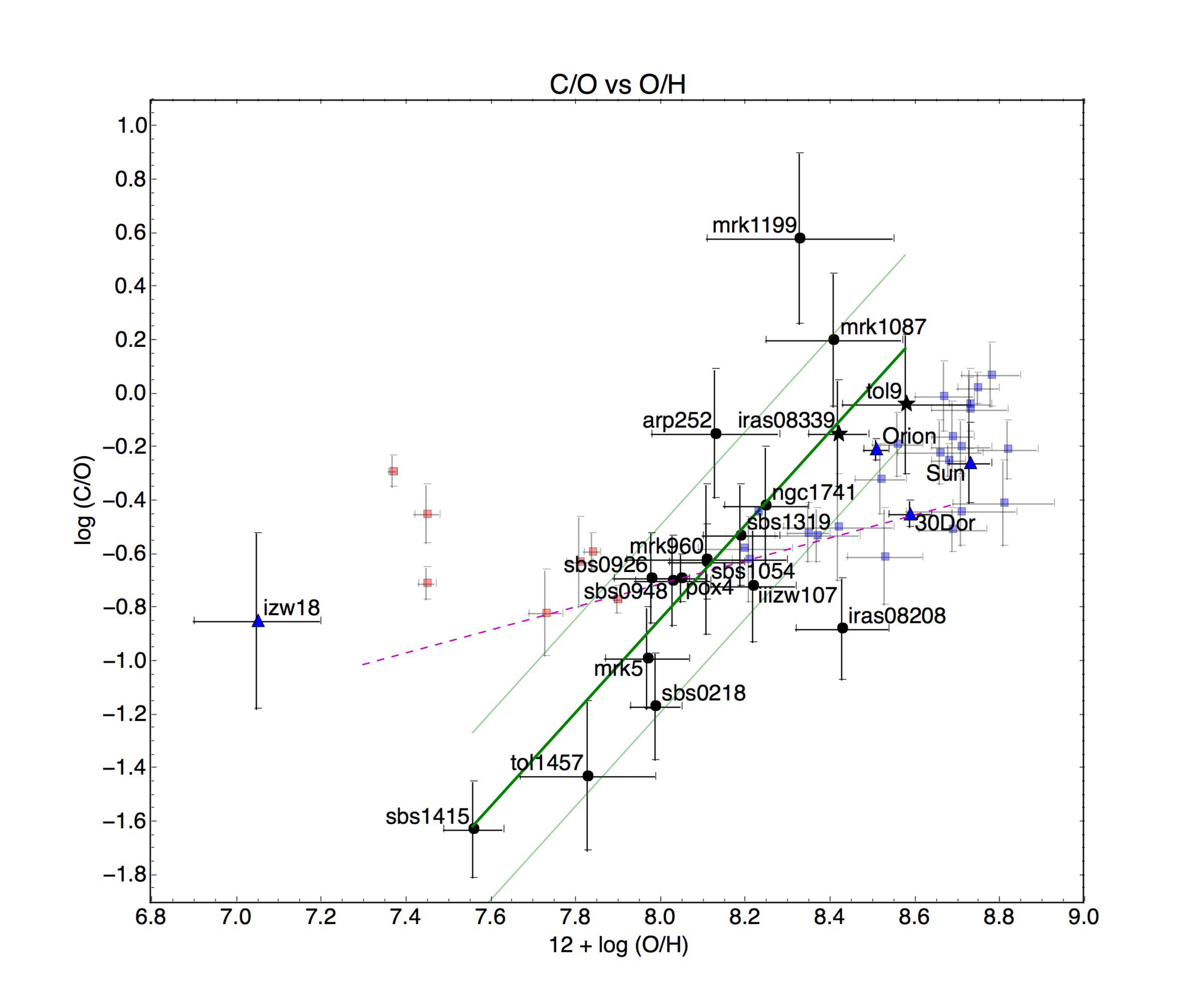}\\
\caption[fig_CO-eps-converted-to.pdf]{
Behavior of C/O versus 12+log(O/H). Filled circles represent the objects for which we obtained a \te\ measurement, the black stars represent those objects for which we used \te\ from the literature, the light red squares represent the data from \citet{ber16}, the light blue squares represent additional data presented in \citet{ber16} taken from \citet{est02, est09, est14}, \citet{gar07}, \citet{lop07}, and the blue triangles represent literature reference points: data for I Zw 18 was taken from \citet{leb13}, data for the Sun was taken from \citet{asp09}, data for 30 Dor was taken from \citet{pei03}, and data for Orion was taken from \citet{est09}. The green line represents a linear fit to our data and the lighter green lines represent the errors of the fit; the dashed magenta line represent the fit proposed in \citet{gar95}.
\label{fCOOH}} 
\end{center}
\end{figure}
\clearpage

\begin{figure}
\begin{center}
\includegraphics[angle=0,scale=0.22]{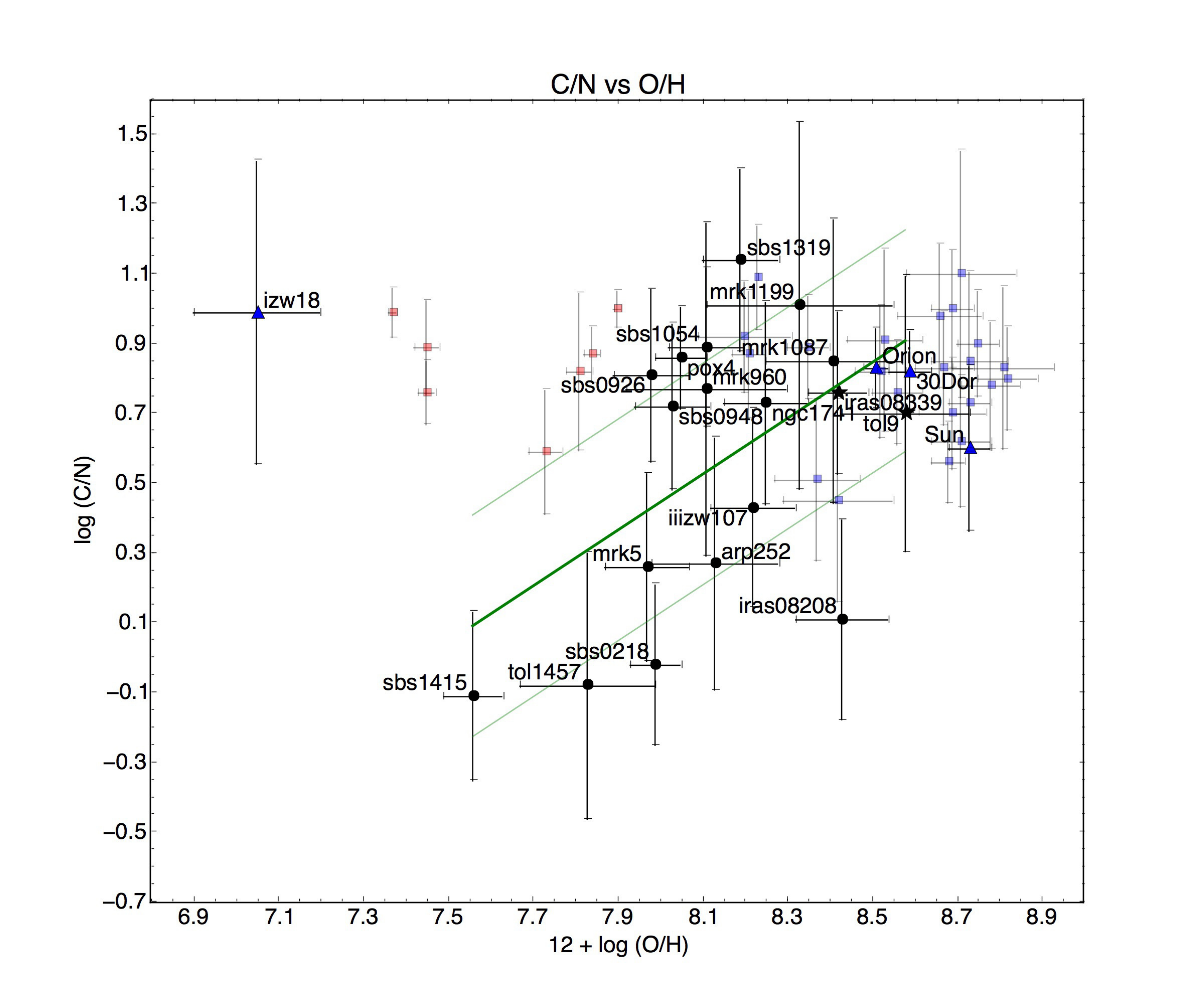}\\
\caption[]{
Behavior of log(C/N) versus 12+log(O/H) as obtained from observations. Symbols are the same as in Figure \ref{fCOOH}. The green line represents a linear fit to our data and the lighter green lines represent the errors of the fit.
\label{fCNOH}} 
\end{center}
\end{figure}
\clearpage

\begin{figure}
\begin{center}
\includegraphics[angle=0,scale=0.22]{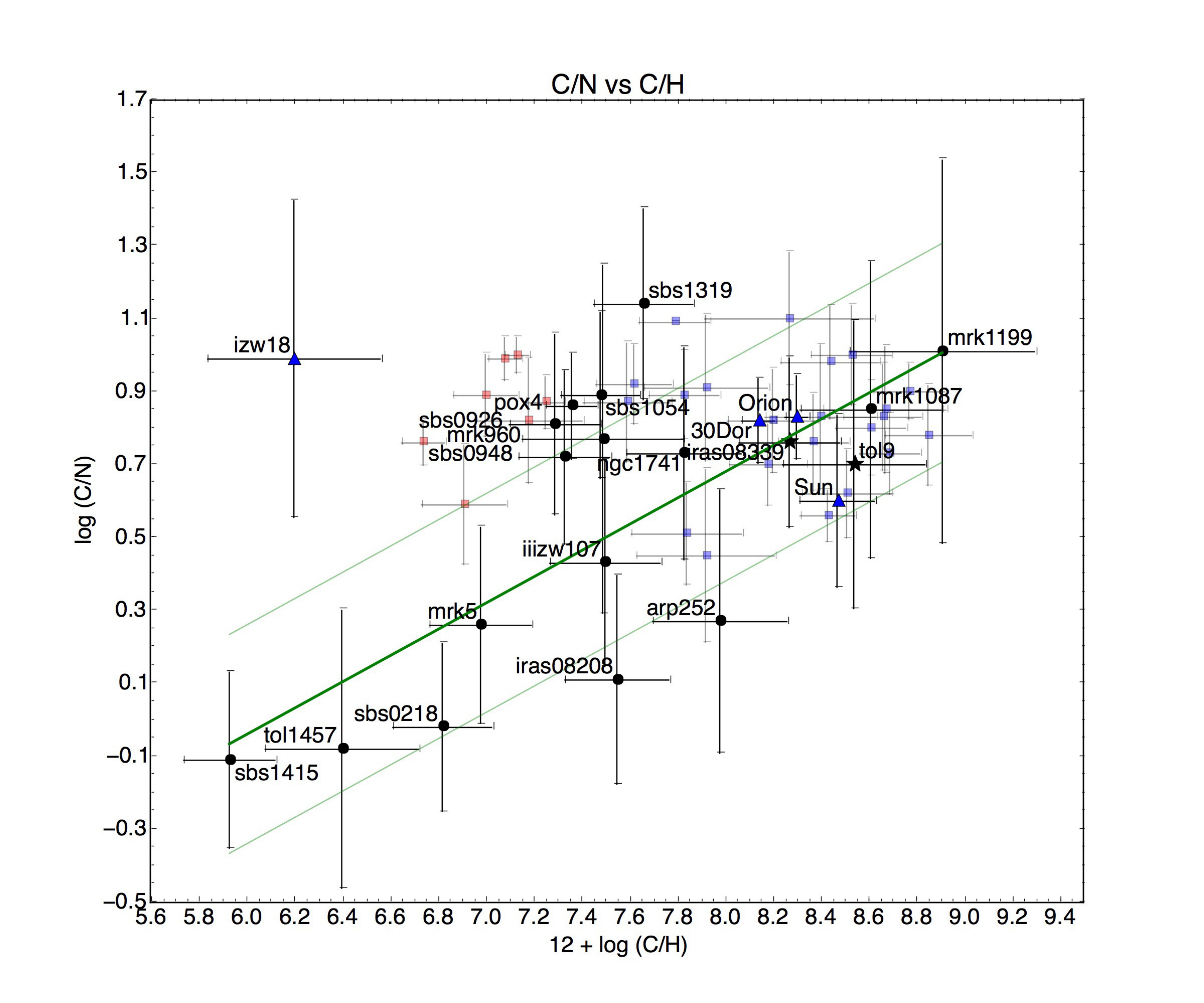}\\
\caption[]{
Behavior of C/N versus 12+log(C/H) as obtained from observations. Symbols are the same as in Figure \ref{fCOOH}. The green line is the best fit found through an MCMC method and the lighter green lines represent the errors of the fit. Only the objects in the sample were used for the fit, the Sun, Orion, 30 Dor, and I Zw 18 are also plotted as reference points. 
\label{fNCCH}}
\end{center}
\end{figure}
\clearpage

\begin{figure}
\begin{center}
\includegraphics[angle=0,scale=0.22]{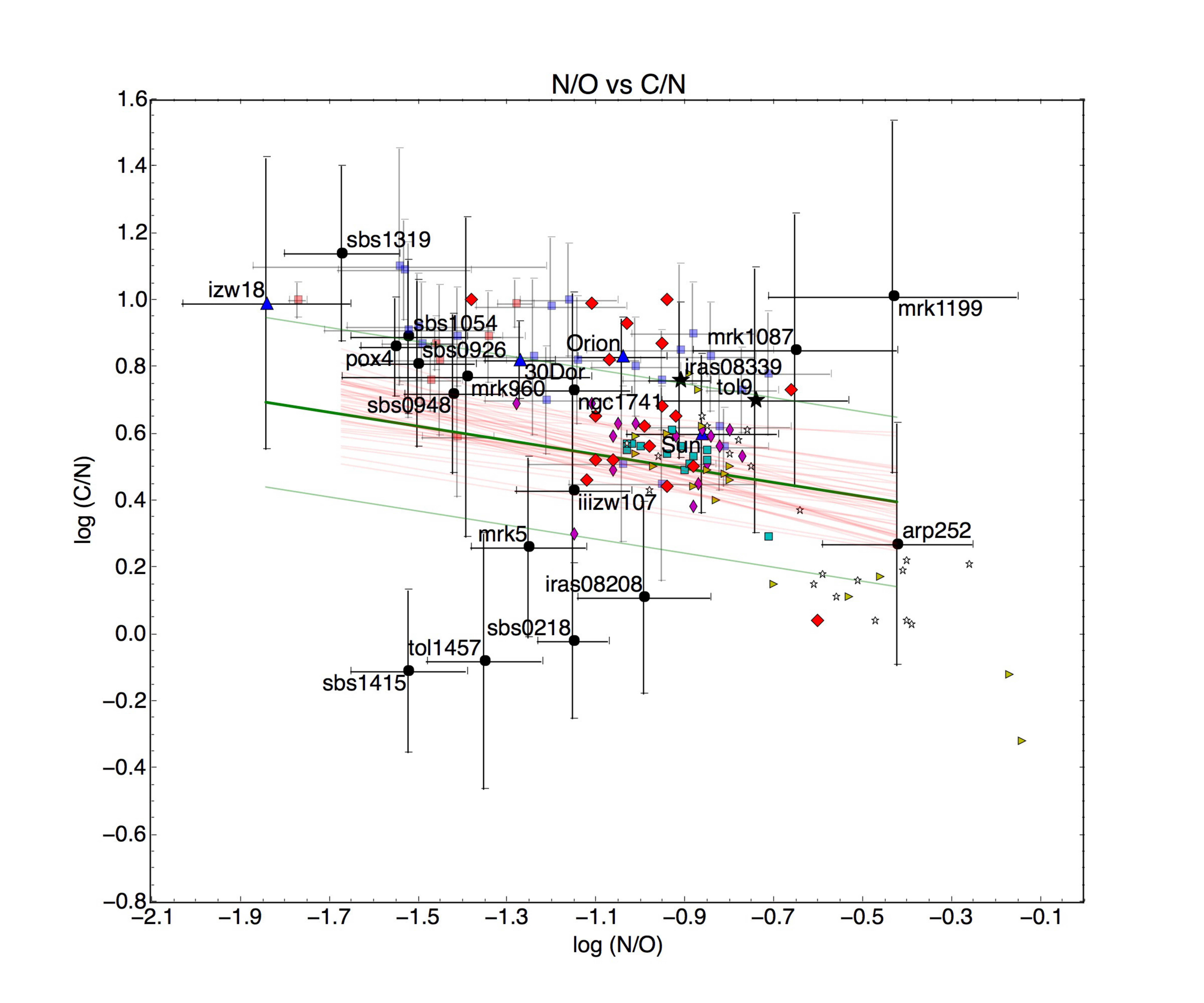}\\
\caption[]{
Behavior of log(C/N) versus log(N/O) as obtained from observations. Filled circles, black stars, and blue triangles represent the same as in Figure \ref{fCOOH}, the light red squares represent the data from \citet{ber16}, the light blue squares represent additional data presented in \citet{ber16} taken from \citet{est02, est09, est14}, \citet{gar07}, and \citet{lop07}. The following are data for disk stars: magenta diamonds from \citet{cun94}, cyan squares from \citet{nie11}, yellow right triangles from \citet{kil92}, red big diamonds from \citet{daf99,daf01a,daf01b}, and white stars from \citet{mor08}. The green line is the best fit found through an MCMC method, while the red lines represent different fits also calculated with an MCMC algorithm. Only the objects in the sample were used for the fit, the Sun, Orion, 30 Dor, and I Zw 18 are also plotted as reference points.
\label{fNOCN}}
\end{center}
\end{figure}
\clearpage

\begin{figure}
\begin{center}
\includegraphics[angle=0,scale=0.18]{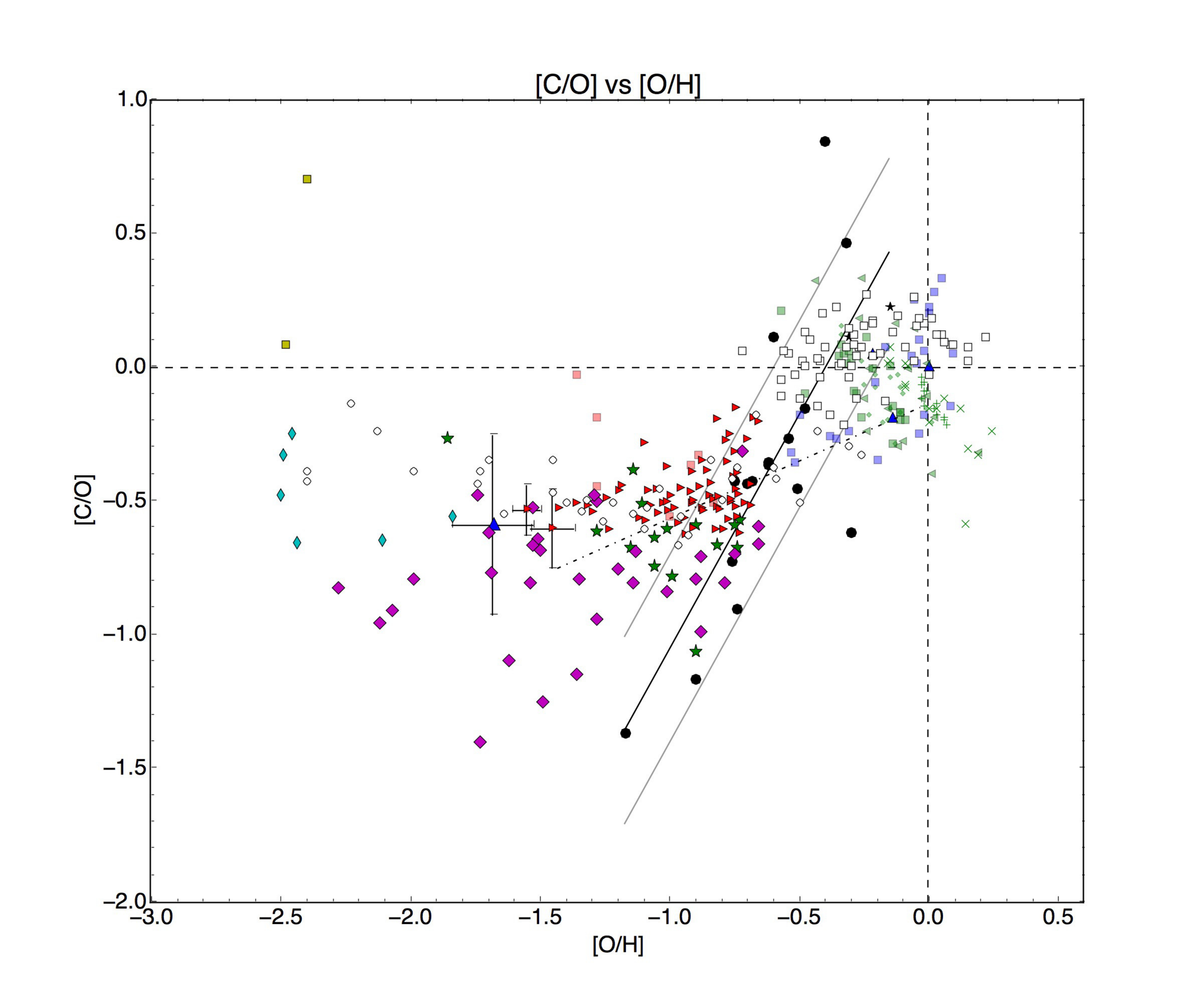}\\
\caption[fig_COcldy-eps-converted-to.pdf]{
Behavior of [C/O] versus [O/H] for various types of objects. Protosolar abundances were taken from \citet{asp09}. %: 12+log(O/H)\solar=$8.73\pm0.05$ and C/O\solar=$-0.26\pm0.07$. 
Filled circles are our STIS data, blue triangles are the same reference points as in Figure \ref{fCOOH}. Light red squares are low-metallicity HII regions of \citet{ber16}, light blue squares are abundances obtained from RLs of \citet{est02, est09, est14}, \citet{gar07}, \citet{lop07}. Cyan thin diamonds are C abundances of \citet{pet08}, yellow squares are very metal-poor DLAs of \citet{coo15}, and light green crosses, plus signs, dots, left triangles, and squares are disk stars of \citet{cun94}, \citet{nie11}, \citet{kil92}, \citet{daf99,daf01a,daf01b}, and \citet{mor08}, respectively. Open smaller circles are Galactic halo stars of \citet{ake04} and white smaller squares are disk stars of \citet{gus99}. The following are approximated with Equation \ref{eqCNNO}: Green stars are extremely low-metallicity SDSS galaxies from \citet{jam15}, red right-side triangles are low-metallicity emission-line galaxy sample of \citet{nav06}, and magenta big diamonds very metal-poor DLAs of \citet{pet08}. The black line represents a linear fit to our data and the grey lines represent the errors of the fit; the black dot-dashed line represents the linear fit of \citet{gar95}. For clarity, error bars are not presented, except for I Zw 18 for which the error bars represent the range of multiple results from the literature.
\label{fCOOHext}}
\end{center}
\end{figure}
\clearpage

\vspace{8mm}
\emph{APPENDIX}.

\section{MCMC Modeling of Photoionized Objects}\label{mcmc}
To verify and further understand the behavior of the chemical abundances in our sample, we ran a series of {\tt Cloudy} models with a {\tt Starburst99} synthetic spectrum as the ionization source through an MCMC method (see Section \ref{mcmc_setup}). We ran about 30,000 models per object letting the chain freely explore the parameter space with the purpose of not creating ad hoc models to our observations. We then restricted \te\ in order to select those models with closer physical conditions, and averaged this subsample to obtain our best estimate for the modeled abundances. The bulk of these values agreed with the abundances of O and C derived from our observations within their uncertainties. Other works, such as \citet{bri04, tre04, per14} and \citet{bla15}, have also used a similar Bayesian analysis to determine the physical conditions and chemical composition of star-forming galaxies. Nonetheless, we have chosen to determine our best estimate of chemical abundances and present our results in a different way. 

\subsubsection{Set Up for the MCMC Photoionization Models}\label{mcmc_setup}
The main goal of the analysis of a problem with an MCMC method is to efficiently sample a set of parameters, using random numbers drawn from a uniform probability in a given range. As a result, probability density functions (PDFs) are obtained for a set of parameters. Such PDFs then show a correlation, anti-correlation, or no relation at all between the set of parameters. Essentially, the MCMC itself consists of random walks that provide a sample of the posterior probability distribution, given certain priors and according to Bayes' law: the posterior probability is proportional to the product of the likelihood and the prior probability,
\begin{equation}
P (\theta, T_e\textnormal{(O)}, n_e | D) \propto \mathcal{L} (\theta, T_e\textnormal{(O)}, n_e )
\;
P(D | \theta, T_e\textnormal{(O)}, n_e ),
\end{equation}
where $\theta$ represents the set of abundance ratios of He/H$\,$ O/H$\,$ C/O, N/O, Ne/O, and S/O, \te (O) represents the electron temperatures of [\ion{O}{2}] and [\ion{O}{3}], $n_e$ represents the electron density, and $D$ the possible true values.

To determine the chemical composition of a set of objects we used 100 walkers and conducted 100 runs per object, which resulted in about 30,000 models of \cldy\ version 13.03 \citep{fer13} per object. We used the Python module {\tt pyCloudy} version 0.8.37 \citep{mor13} to include \cldy\ into our code. The priors were set so that the trial abundances (or random walks) were physically plausible values for the set of abundances (or dimensions). The ionizing source we used for the \cldy\ models was a synthetic spectrum generated with {\tt Starburst99} version 7.0.1 \citep{lei14} at an age of 10$^8$ years. We used a continuous star formation rate (SFR) for the {\tt Starburst99} spectra following \citet{gol97}, who find that an instantaneous SF burst provides an unrealistic short range of ages for their sample of luminous infrared galaxies, and \citet{kew01}, who find that a continuous SFR  agrees better with their sample of infrared starburst galaxies, whose spectrum is dominated by emission from W-R stars. Additional conditions used for the {\tt Starburst99} spectra were: a Kroupa initial mass function (IMF) with two exponents  of 1.3 and 2.3 \citet{kro14} and mass boundaries of 0.1, 0.5, and 120 M\solar; the 2012/13 Geneva stellar evolutionary tracks with rotation velocities of 40\%\ of the break-up velocity and $Z=0.014$ \citep{eks12, geo13}. All other parameters were set to the default values in {\tt Starburst99} (see http://www.stsci.edu/science/starburst99/docs/code.html).

\subsubsection{MCMC Photoionization Models Input and Comparison to Observations}
The input trial abundances for the \cldy\ models were varied according to the MCMC algorithm in the Python module {\tt emcee} \citep{for13}. The \cldy\ output contains line intensities as well as temperatures of [\ion{O}{3}] and [\ion{O}{2}]. We assumed that the uncertainties of the observations follow a gaussian distribution, therefore we used $\chi^2$ as a likelihood function -more specifically we used $ln(-\chi^2/2$). 

We calculated $\chi^2$ from comparing the observed line intensities to the modeled ones. We used the intensity relative to H$\beta$ of 9 lines to determine a $\chi^2$ value per model: \ion{H}{1} 4340, 4861, and 6563, \ion{He}{1} 5876, \ion{He}{2} 4686, [\ion{O}{2}] 3727, [\ion{O}{3}] 5007, \ion{C}{3}] 1909, and [\ion{S}{3}] 9532 \AA. For the \cldy\ models we used the following options: electron density of 150 cm$^{-3}$, cosmic ray background, atom H-like levels, atom He-like levels, no fine opacities, no molecules, and no level 2 lines. 

An initial chemical composition was also given as input for each model. One of the advantages of the MCMC method is that the initial guess is does not have an effect on the ``final" answer because the set of walkers obtained at the end should be independent samples from the distribution \citep[for a more detailed discussion on the initial guess see][]{for13}. We initiated the chains with random abundance values in the following ranges of typical observed values {for \ion{H}{2} regions}: $9.8\leq$He$\leq10.0$, $8.3\leq$O$\leq8.6$, $-0.3\leq$C/O$\leq0.1$, $-1.3\leq$N/O$\leq-1.0$, $-0.3\leq$Ne/O$\leq0.0$, and $-2.0\leq$S/O$\leq-1.7$, in units of 12+log(X/H) and log(X/O), respectively. The initial abundances were varied according to the MCMC algorithm. These were checked against the prior probability (allowed values): $9.5\leq$He$\leq12.0$, $7.5\leq$O$\leq8.7$, $-1.6\leq$C/O$\leq1.7$, $-1.7\leq$N/O$\leq-0.4$, $-1.0\leq$Ne/O$\leq0.01$, and $-2.3\leq$S/O$\leq-1.4$. We used a top hat function for each of the abundances:
\begin{equation}
f(x) = \left\{ \begin{array}{lcl}
			- ln(b-a) && a<x<b \\
	               -\infty        && else
	           \end{array}
          \right.
\end{equation}
where $a$ and $b$ are the abundance values corresponding to allowed minimum and maximum, respectively. If the state is accepted, then the values are converted to abundances relative to hydrogen in order to be sent as input for the \cldy\ models. With the mentioned options each model took between 3 and 5 minutes to run.

\subsection{Results from MCMC Photoionization Modeling}
The result of the MCMC is a distribution of probabilities of occurrence for a set of parameters. Figure \ref{fmcmcPOX4} (and the online only Figure set) shows the relations between our set of parameters: He/H, O/H, C/O, N/O, Ne/O, and S/O, in units of 12+log(X/H) or log(X/O). The blue crosses represent the benchmark abundances (i.e. abundances obtained from our STIS observations), and the red crosses represent the best estimate from the models. There are various methods to obtain a value from this distribution in order to compare it with the measured abundances from observations. One of such methods is to maximize of the likelihood and use that value (particularly useful when the histogram presents a gaussian distribution, see Figure \ref{fmcmcPOX4}). However, the major advantage of the MCMC analysis is the availability of a distribution of values. The average or median values use such distribution to obtain a best estimate. In this work we used the average values; the median values yield essentially the same numbers. 

In order to account for the observed oxygen temperatures, we made a sub-sample of models taking only those with temperatures of [\ion{O}{3}] and [\ion{O}{2}] $T\textsubscript{model}\leq$T\textsubscript{observed}$\pm2500\;$K, and we took the average of this sub-sample. {We chose an arbitrarily large range of temperatures in order to encompass all models within the range of uncertainties of the observations.} Table \ref{modchem_comp} presents the gaseous abundances in units of 12+log(X/H) or log(X/O) of He/H, O/H, C/H, N/H, Ne/H, and S/H as obtained from the MCMC Photoionization Modeling method for the same elements. {The uncertainties we present} for the modeled abundances were obtained from the 25 and 75 percentiles of the subsample of models (i.e. all models with $T$\textsubscript{observed}$\pm2500$ K). These uncertainties are large due to our decision of letting the chain explore the parameter space as freely as possible. For reference, we created four spectral zoom-in windows that show the major emission lines per object. Figure \ref{pox4sup} is an example of these zoom-in windows (the corresponding figures for all other objects are presented only in the online version).

\subsection{Analysis of the MCMC Photoionization Models}\label{anamcmc}
\begin{figure}
   \begin{center}
   \includegraphics[angle=0,scale=0.35]{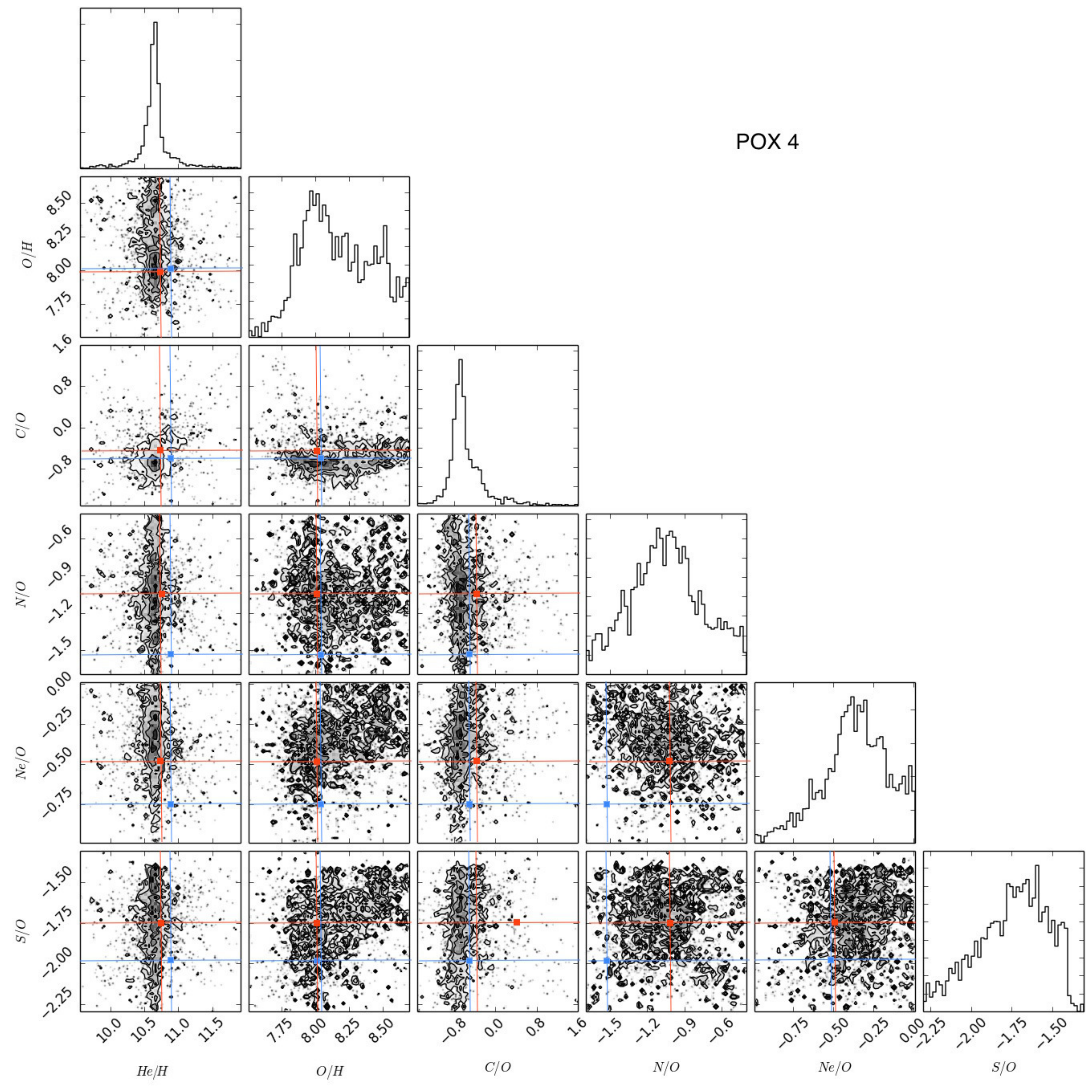}\\
   \caption[fmcmcPOX4-eps-converted-to.pdf]{
Behavior of the probability of occurrences for the set of abundances of 12+log(He/H), 12+log(O/H), log(C/O), log(N/O), log(Ne/O), and log(S/O) according to our MCMC photoionization modeling for POX 4. The blue crosshairs represent the measurements from the observations and the red crosses represent our best estimate from the models. The uppermost panel in each column is showing the histogram of the variable directly below. The figure was created with \texttt{corner.py} \citep{corner.py}.
   The complete figure set (18 images) is available in the online journal.
   \label{fmcmcPOX4}}
   \end{center}
   \end{figure}

\begin{figure}
   \begin{center}
   \includegraphics[angle=0,scale=0.2]{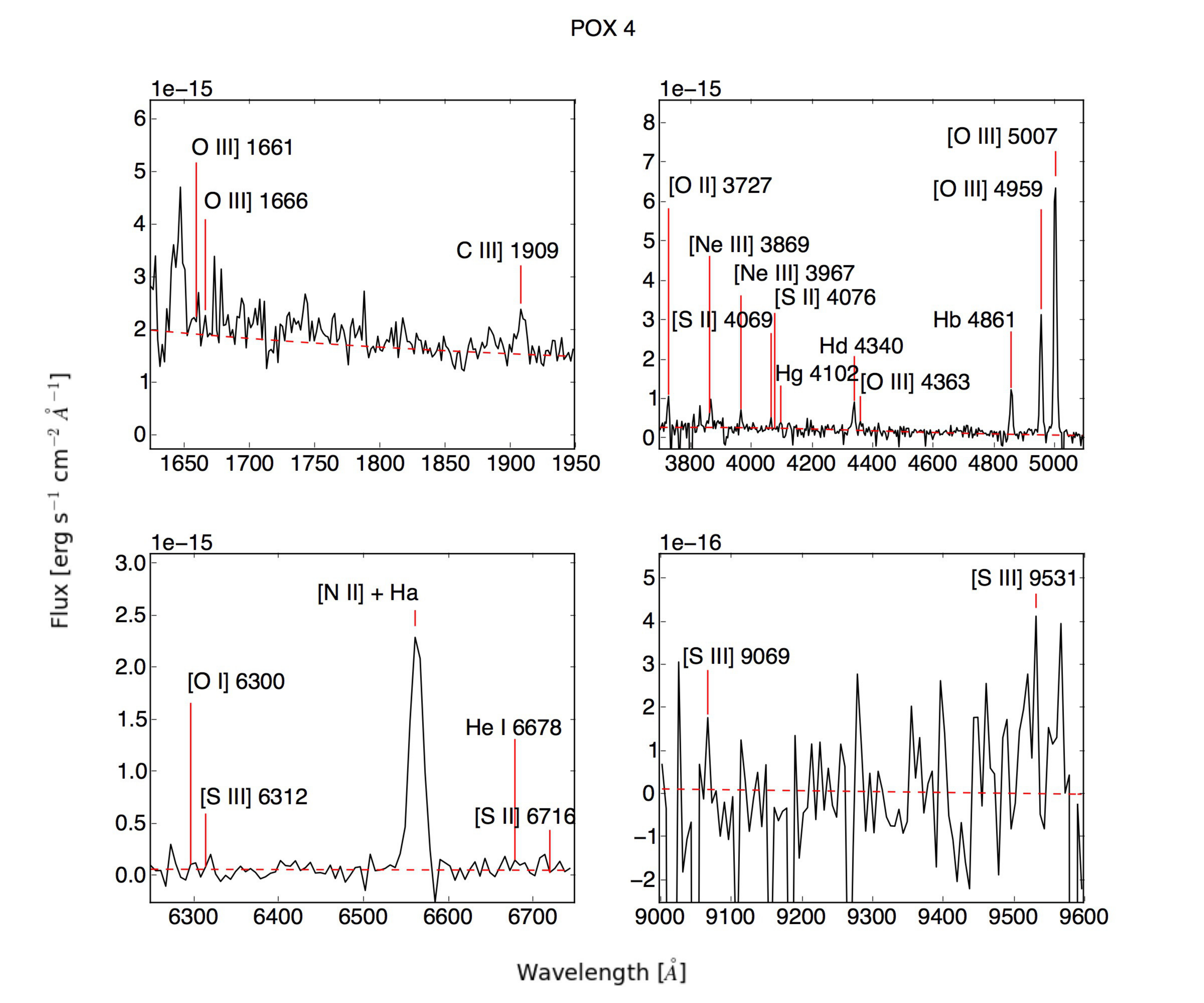}\\
   \caption[]{
   Zoom-in of the spectrum for the UV, blue, red, and IR regions for POX 4.
   The complete figure set (18 images) is available in the online journal.
   \label{pox4sup}}
   \end{center}
   \end{figure}

\begin{figure}
\begin{center}
\includegraphics[angle=0,scale=0.44]{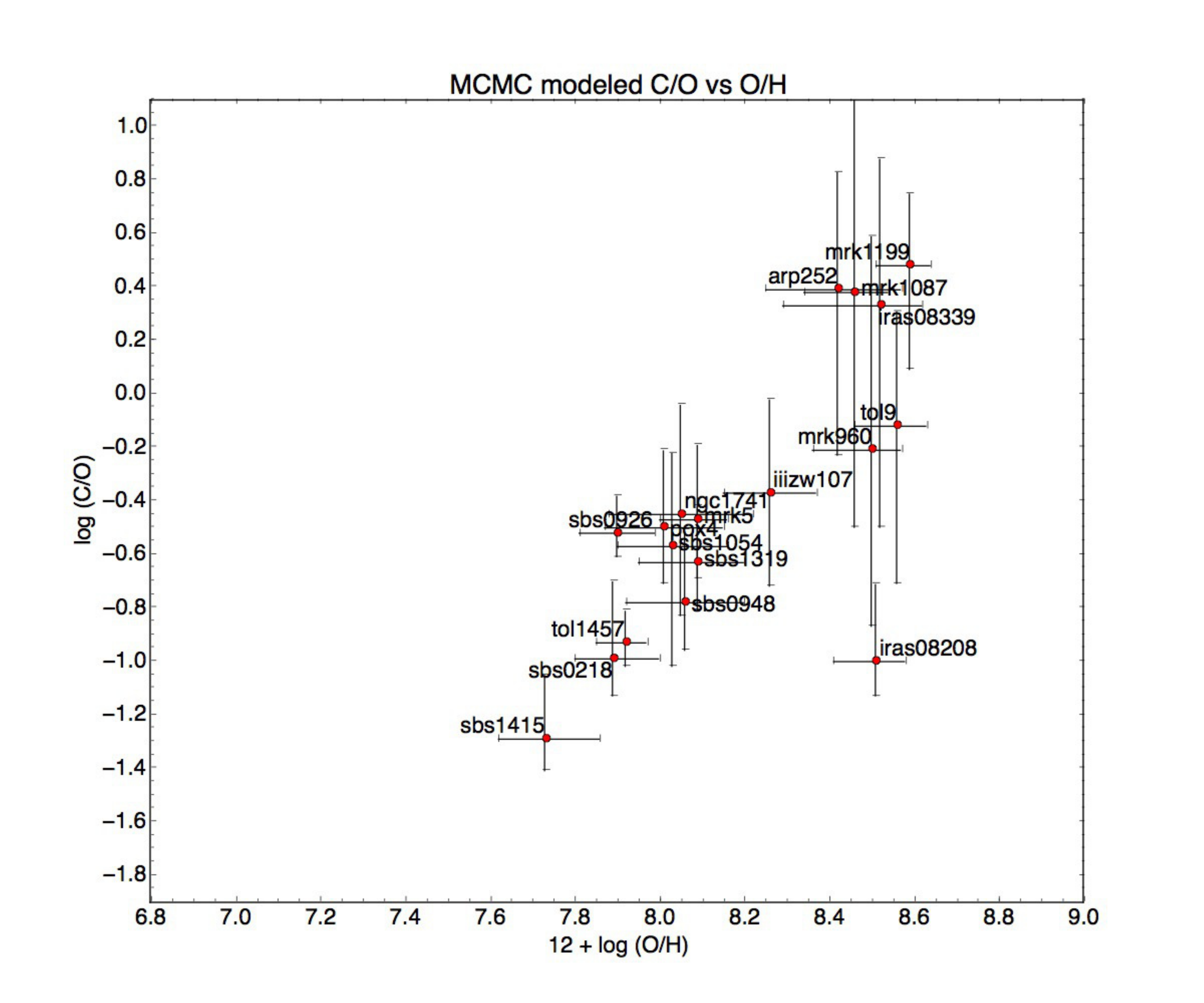}\\
\caption[fig_COcldy-eps-converted-to.pdf]{
Behavior of C/O versus 12+log(O/H) as obtained from MCMC. Symbols are the same as in Figure \ref{fCOOH}. 
\label{fCOOHcldy}}
\end{center}
\end{figure}

The C/O ratios as obtained from the MCMC photoionization models are plotted against the gaseous modeled oxygen abundance in Figure \ref{fCOOHcldy}. We see a similar behavior as that shown in Figure \ref{fCOOH}. {The uncertainties are large due to the freedom with which the MCMC was run. We did this with the purpose of exploring most of the parameter space. Figure \ref{fmcmcPOX4} shows an example of the} PDF relations between the set of parameters for each object. {For POX 4, we find that there are three instances with no correlation: (i) panel of C/O vs. He, which shows a circular contour, (ii) panels like S/O vs. Ne/O, which show points in the entire space, and (iii) panels like S/O vs. C/O, in which one of the parameters has a constant value. On the other hand, panels like Ne/O vs. O show a potential linear correlation.} In this figure we see no anti-correlations.

\begin{figure}
\begin{center}
\includegraphics[angle=0,scale=0.22]{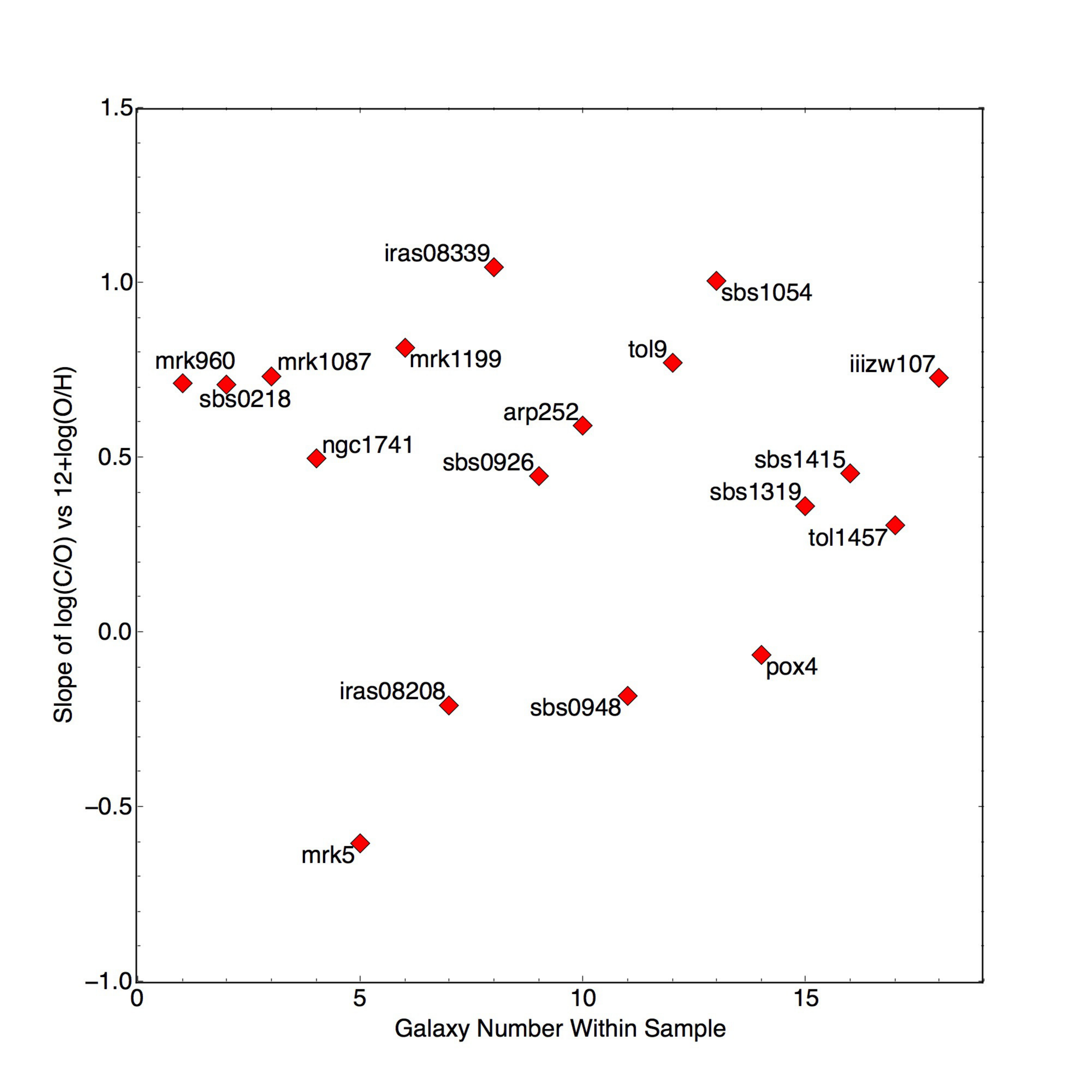}\\
\caption[]{
Behavior of the slopes of log(C/O) to 12+log(O/H) as obtained from the MCMC photoionization modeling per galaxy.
\label{COslopes}}
\end{center}
\end{figure}

\begin{figure}
\begin{center}
\includegraphics[angle=0,scale=0.7]{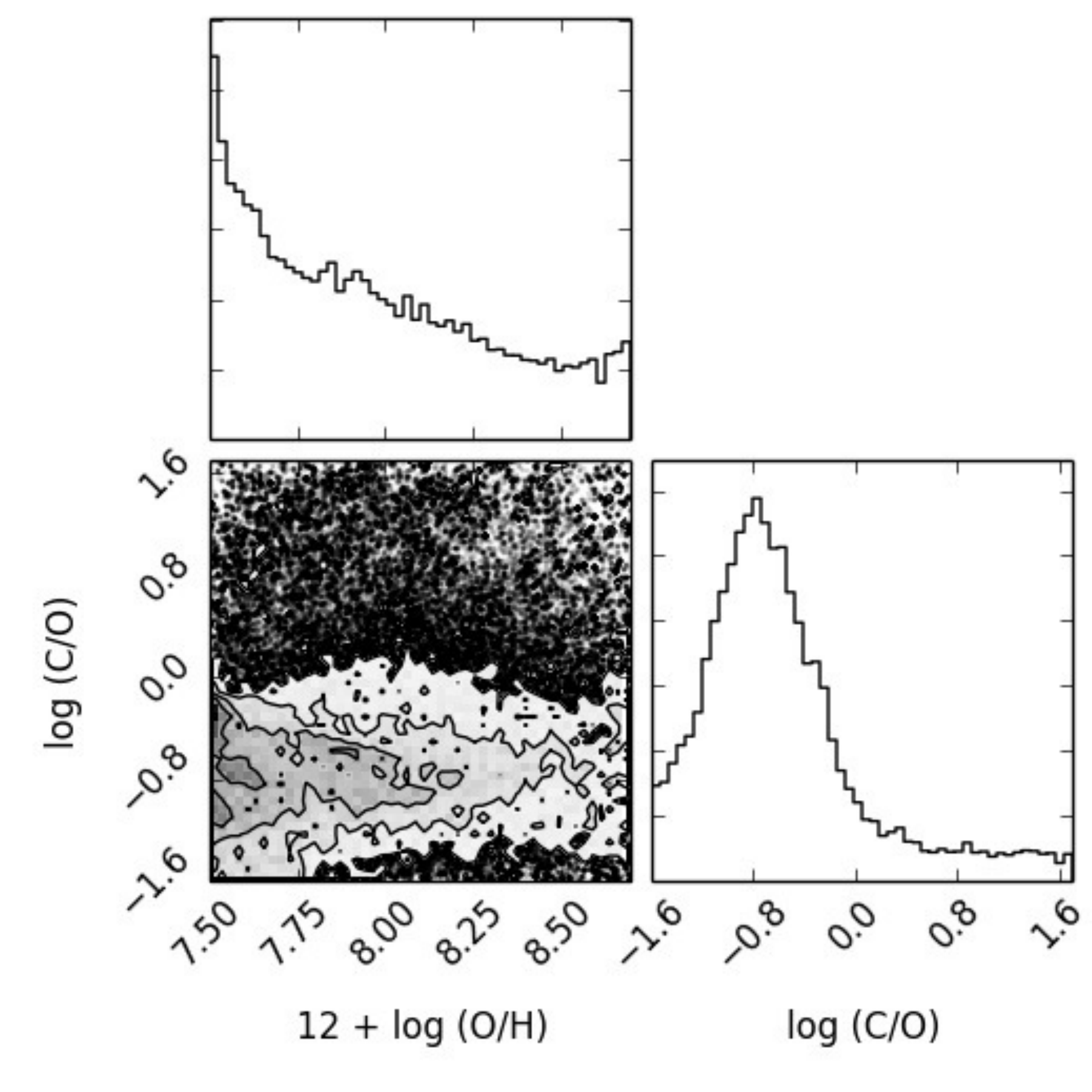}\\
\caption[fwholeCO-eps-converted-to.pdf]{
Behavior of probability of occurrences for the set of abundances log(C/O) versus 12+log(O/H) for the whole sample. The upper most panel and the right most panel are showing the histogram of the variable directly below. This figure includes all the \cldy\ models for all the objects in the sample.
\label{fwholeCO}}
\end{center}
\end{figure}

{From our MCMC models (i.e. the PDF diagrams like Figure \ref{fmcmcPOX4}), we} observed that for the log(C/O) versus 12+log(O/H) diagram about half of the objects in our sample appear show an increasing slope whereas the other half appear to show a somewhat flat behavior. We determined the slopes of this diagram for the models ran per galaxy, and found that the range of slopes is -0.6 to 1.0, with an average slope of 0.4 and a median slope of 0.5. Figure \ref{COslopes} shows the slopes from our MCMC photoionization models per galaxy. {Each} slope was determined from a linear fit to all the {models for that object} in the log(C/O) versus 12+log(O/H) diagram. We observe that these slopes do not seem to be related to metallicity, distance, redshift, carbon abundance, type of object, interaction degree, S/N of the pertinent lines, number of W-R stars versus O stars, or ionization degree. To check if this was a random effect of the MCMC, we re-ran the chain three times with the same conditions for three objects (the least and most metallic objects, and an intermediate-metallicity object: SBS 1415$+$437, Tol 9, and SBS 0218$+$003, respectively). The behavior of the slope remained the same for all three objects. We then decided to combine all the models for all the objects into a single figure of C/O versus 12+log(O/H). The result is presented in Figure \ref{fwholeCO}, which shows a clear flat behavior with main or most probable value of C/O of about $-0.8$. This flat behavior in the context of the MCMC analysis suggests that there is no obvious correlation between log(C/O) and 12+log(O/H). Nonetheless the value of C/O seems to be well restricted to values between -1.6 and 0.0. The bulk of the sample in this paper has C/O values within this range. {From the analysis of Figures \ref{fCOOH}, \ref{fCNOH}, and \ref{fNCCH}, there seems to be a specific commonality that separates points from our sample from literature references, and we suspect that could be the top-heavy IMF that all objects in our sample share. If this is true, the slope behavior we observe in the models is to be expected since the IMF was the same for all models. A separate study would be required to asses if this is indeed the case.} 

Using all the models for all the objects, we also studied log(C/N) versus 12+log(O/H), and log(N/O) versus log(C/N). We find that both figures confirm the clear linear correlation found with our observed data for both cases. Interestingly, the main value of [C/O] for metal-poor halo stars according to \citet{tom92} is about $-0.60$ and about $-0.50$ according to \citet{fab09}, which translates into C/O of $-0.9$ and $-0.8$, respectively, adopting the protosolar values for C/O\solar$\,=-0.26\pm0.07$ and 12+log(O/H)\solar=$8.73\pm0.05$ \citep{asp09}. \citet{est14} determined carbon abundances for a small sample of objects via RLs. They also compared their C abundances to Galactic halo stars. Their C measurements agree with the average C/O from Galactic halo stars shown in Figure 6 of \citet{est14}, also between $-0.9$ and $-0.8$. {If our hypothesis of the IMF playing an important role is true, then these plots would again imply that there is a specific behavior for objects that have a similar IMF, and that halo stars are well described by a Kroupa IMF like the one we used for generating the models.}

Since intermediate-mass stars contribute to the production of C, the ejection of carbon is delayed with respect to oxygen. Hence, the C/O ratio could be used as an indicator of the relative age of stellar systems \citep{gar95}. Figure 8 in \citet{boy13}, which presents the carbon-rich to oxygen-rich (or metal-rich, we will refer to this ratio as C/M) star ratio in the age-metallicity plane for thermally pulsing AGB (TP-AGB) models of \citet{mar13}, shows a striking similarity to Figure \ref{fwholeCO} in the present paper. In this figure we present the behavior of log(C/O) versus 12+log(O/H) for all the \cldy\ models for the whole sample. If the similarity between these figures is not coincidental, the blue part in the \citet{boy13} figure (corresponding to stars with C/O$<$1) would indicate that the maximum C/M star ratio corresponds to a more positive value of C/O because AGB stars have reached the third dredge-up process and are expelling their carbon into the ISM. As metallicity decreases, it is more difficult to form massive stars, hence there will be more sources of carbon-rich stars. 

To test if the effect of a different ionizing spectrum on the MCMC \cldy\ models, we generated four different synthetic ionizing spectra and re-ran the MCMC for the least and most metallic objects in our sample: SBS 1415$+$437 and Tol 9, respectively. The four {\tt Starburst99} ionizing spectra had constant SFR and were taken at an age of $10^8$ years. The four models had the following: (i) Geneva stellar evolutionary tracks with rotation velocities of 40\%\ of the break-up velocity and $Z=0.014$, (ii) same Geneva stellar evolutionary tracks with $Z=0.001$, (iii) Geneva stellar evolutionary tracks with rotation velocities of 0\%\ of the break-up velocity and $Z=0.014$, and (iv) same Geneva stellar evolutionary tracks with $Z=0.001$. All other initial {\tt Starburst99} conditions were left as described in Section \ref{met}. Spectrum (i) was the same used to ran all models for all objects in the sample. We simply re-ran the chain with the same conditions to see how much the results varied. In order to obtain final abundance values from the MCMC, we used the same average method as previously described in Section \ref{met}. We find that there is little variation in the best estimated abundances and the abundances from the models between the $25^{th}$ and $75^{th}$ percentiles for all four combination of stellar tracks with metallicity.

{As a control experiment to determine how accurate is the MCMC method we used for abundance-verifying,} we used the line ratios of NGC$\,$5253 presented in \citet{kob97} to run a chain for this object. In this case we used the abundances given in \citet{kob97} for the \ion{H}{2} region as initial guess in units of 12+log(X/H): He$=10.90\pm0.05$, O$=8.19\pm0.07$, C$>7.46$, N$=7.33\pm0.13$, Ne$=7.32\pm0.16$, and S$=6.82\pm0.17$. The O$^{++}$ and O$^+$ temperatures they derived are 11,250 and 11,850 K, respectively. Just like with our sample, we initiated the walkers with random positions in the range of typical observed values. The MCMC diagrams for NGC$\,$5253 show that the models did not converge, however, our sample selection method still worked and we were able to obtain {a} first approximation {to the abundances}. Just as {with our sample,} we used only the models with $T\textsubscript{model}\leq$T\textsubscript{observed}$\pm2500$ K, and we took an average of {these. We determined the uncertainties from the} 25$^{th}$ and 75$^{th}$ percentiles, and since the chain did not converge, {we expected these uncertainties to be} large. This yielded the following abundances in units of 12+log(X/H): He/H$=10.74\pm^{0.59}_{0.65}$, O/H$=8.04\pm0.23$, C/H=$8.00\pm^{0.79}_{0.87}$, N/H$=7.03\pm0.28$, Ne/H$=7.63\pm0.23$, and S/H$=6.32\pm0.25$. For those object where there is no $a\;priori$ knowledge of the oxygen temperatures, a solution would be to carefully fit a \cldy\ model to obtain an approximate of the high and low ionization zone temperatures and use them as a proxy for T\textsubscript{observed}. We then used these center values as initial condition to re-run the chain. We initiated the walkers with random values but within a tighter ``ball" around them (i.e. center value $\pm$0.2 dex) and ran a twice as long chain (200 runs with 100 walkers). Though the chain did not converge again, the resulting abundances for this re-run were the following: He/H$=10.78\pm{0.07}$, O/H$=8.09\pm0.08$, C/H=$8.06\pm{0.12}$, N/H$=7.06\pm0.12$, Ne/H$=7.68\pm0.11$, and S/H$=6.39\pm0.10$, also in units of 12+log(X/H). These results suggest that our sub-sample selection method is able to narrow down the set of abundance parameters {to obtain a crude first approximation to the abundances} even when the chain does not converge, as long as there is an $a\;priori$ knowledge of the high- and low-ionization zone temperatures. 

   \clearpage

   \clearpage

\clearpage

\clearpage

\clearpage

\end{document}